\pdfoutput=1 

\documentclass[3p,authoryear]{elsarticle}

\usepackage{ecrc}
\usepackage{lineno,hyperref}
\usepackage{graphicx} 
\usepackage{subfigure} 
\usepackage{epstopdf}
\usepackage{amsmath} 
\usepackage{amsthm} 
\usepackage{amsfonts} 
\usepackage{amssymb} 
\usepackage{algorithm} 
\usepackage{algorithmic}
\usepackage{url}
\usepackage{float} 
\usepackage{ulem} 
\usepackage{booktabs}
\usepackage{color}

\volume{00}

\firstpage{1}

\journalname{Advances in Space Research}

\runauth{S. Lyu et al.}

\jid{ASR}

\jnltitlelogo{Advances in Space Research}

\usepackage{amssymb}

\usepackage{lineno}

\usepackage[figuresright]{rotating}

\begin{document}
	
	\begin{frontmatter}
		
		
		\title{Optimal Stereoscopic Angle for Reconstructing Solar Wind Inhomogeneous Structures}
		\author[address1,address2]{Shaoyu Lyu}
		\ead{lsy1997@mail.ustc.edu.cn}
		
		\author[address1,address2]{Xiaolei Li\corref{cor1}}
		\ead{lxllxl@mail.ustc.edu.cn}
		
		\author[address1,address2,address3]{Yuming Wang\corref{cor1}}
		\ead{ymwang@ustc.edu.cn}
		
		\cortext[cor1]{Corresponding author.}
		\address[address1]{CAS Key Laboratory of Geospace Environment, Department of Geophysics and Planetary Sciences,\\University of Science and Technology of China(USTC), Hefei 230026, China}
		\address[address2]{CAS Center for Excellence in Comparative Planetology, USTC, Hefei 230026, China}
		\address[address3]{Mengcheng National Geophysical Observatory, USTC, Hefei 230026, China}
		
		
		\dochead{}
		
		
		
		
		
		\begin{abstract}
			This paper is aimed at finding the best separation angle between spacecraft for the three-dimensional reconstruction of solar-wind inhomogeneous structures by the CORrelation-Aided Reconstruction(CORAR) method. The analysis is based on the dual-point heliospheric observations from the STEREO HI-1 cameras. We produced synthetic HI-1 white-light images containing artificial blob-like structures in different positions in the common field of view of the two HI-1 cameras and reconstruct the structures with CORAR method. The distributions of performance levels of the reconstruction for spacecraft separation of $60^{\circ}$, $90^{\circ}$, $120^{\circ}$ and $150^{\circ}$ are obtained. It is found that when the separation angle is $120^{\circ}$, the performance of the reconstruction is the best and the separation angle of $90^{\circ}$ is the next. A brief discussion of the results are given as well. Based on this study, we suggest the optimal layout scheme of the recently proposed Solar Ring mission, which is designed to routinely observe the Sun and the inner heliosphere from multiple perspectives in the ecliptic plane.
		\end{abstract}
		
		\begin{keyword}
			Solar Ring \sep Solar wind transients \sep Stereoscopic observations \sep 3D reconstruction
		\end{keyword}
		
	\end{frontmatter}
	
	
	\section{Introduction}
	

	The study of the evolution of solar wind structures from the solar surface to the interplanetary space is an important issue in space physics, which involves the dynamic mechanism of plasma in the solar-terrestrial space and helps to forecast space weather in the geospace. With the development of white-light imaging instruments, the solar wind, especially the inhomogeneous components like coronal mass ejections (CMEs) and 'blobs', can be observed. CMEs are typically large-scale solar eruption events delivering a large amount of energy and matter into interplanetary space, and may trigger magnetospheric disturbances affecting human activities\citep{Siscoe2006}. 'Blobs' are small-scale transients generally originating from the cusps of helmet streamers and propagating outward with slow solar wind\citep{Wang1998,Sheeley1999,Sheeley2009}, or from the boundaries of weak coronal holes\citep{Lopez-Portela2018}.

	In spite of the high-resolution imaging data from space-based instruments like the Large Angle and Spectrometric Coronagraph (LASCO,\citealp{Brueckner1995}) on board Solar and Heliospheric Observatory (SOHO,\citealp{DOMINGO1995}), the single-viewpoint observation provides only two-dimensional (2D) information of solar wind structures on the plane of sky (POS) and thus has unavoidable projection effect for three-dimensional (3D) reconstruction. For complex structures like CMEs, the kinematic properties of real and projected objects can be significantly different\citep{Howard2008a}. It is revealed that the projection effect is prominent for most full-halo CMEs with a slow projected speed within $45^{\circ}$ from the Sun-Earth line\citep{Shen2013}. Some reconstruction methods based on single-viewpoint observations were developed. According to geometric principles, some methods such as 'Point P'\citep{Howard2006}, 'Fixed-$\phi$'\citep{Kahler2007} and 'Harmonic Mean'\citep{Lugaz2009,Wang2013,Rollett2013} can make an estimate of the location and propagating direction of the dynamic structures. Furthermore, given the morphological assumptions of certain solar transients like cone model\citep{Zhao2002,Xie2004,Xue2005,Zhao2008} or self-similar expansion model\citep{Davies2012} for radial-propagated CMEs or their fronts, the simple 3D geometrical model is iterated to get a best fit for the observation data, which is called Forward Modelling (FM) method. These methods are suitable for reconstruction of temporal solar-wind structures. With the polarization distribution of Thomson-scattering light, \textcolor[rgb]{0,0,0}{polarimetric reconstruction technique\citep{JACKSON1995,Moran2004,Dere2005,Moran2010,Pagano2015} can be used to locate the centroid of corona along the line of sight (LOS), and \citet{Lu2017} improved this method by removing CME-irrelevant structures.} 
	
	\textcolor[rgb]{0,0,0}{The successful launch of the two Solar Terrestrial Relations Observatory (STEREO, \citealp{Kaiser2008}) spacecraft enables simultaneous observations from multiple vantage points in the heliosphere.} The combination of imaging data from different viewpoints provides 3D information of solar wind structures. With a multi-view dataset, more sophisticated models can be applied for FM method such as the Graduated Cylindrical Shell (GCS) model\citep{Thernisien2006,Thernisien2009,Mierla2009,Mierla2010}, and therefore provides more convincing results. Assuming a constant propagating velocity, the prominent tracks on the time-elongation 'J-Map'\citep{Sheeley1999a,Davies2009} from stereoscopic white-light observations can locate the 3D propagation paths of small transients such as blobs to determine their directions and radial velocities\citep{Sheeley2009,Sheeley2010,Liu2010a}. Associated with geometrical principles, the Geometric Localization method\citep{Pizzo2004,Koning2009} is a direct application of triangulation to reproduce the edges of objects. Extending to three-point observation, the Mask Fitting method\textcolor[rgb]{0,0,0}{\citep{Feng2012,Feng2013}} \textcolor[rgb]{0,0,0}{is proposed for reconstructing solar structures} with best-fitting projection on images from three different viewpoints. As to correlation analysis, Local Correlation Tracking (LCT) method\citep{Mierla2009,Mierla2010,Feng2013} is one important application that finds the analogical tendency of spatial variation on dual-view images representing the same 3D structure. Similar to LCT method, CORrelation-Aided Reconstruction (CORAR) method is developed for detection and localization of solar-wind inhomogeneous structures from dual perspectives of STEREO-A and B, and is proofed to be efficient for 3D reconstruction of solar transients\citep{Li2018,Li2019}.
	
\begin{figure*}[htb]
		\centering
		\includegraphics[width=0.75\columnwidth,height=0.7\columnwidth]{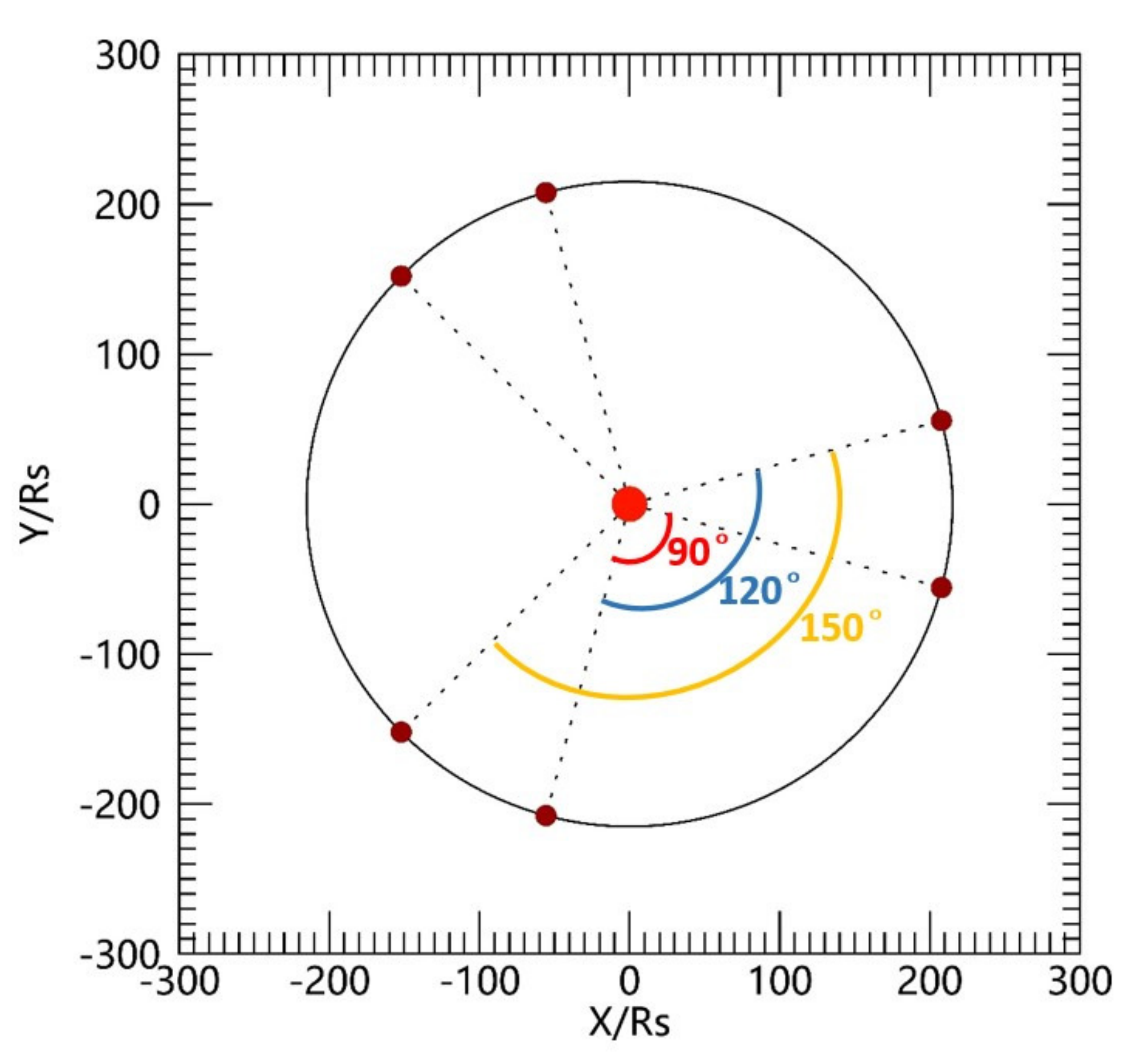}
		\caption{The spacecraft scheme for the Solar Ring mission. Possible separation angles of two observers are labelled with different colors, and the circle represents the orbit of six spacecraft located with brown points on the circle.} 
		\label{fig:2-1}
	\end{figure*}

	Based on the potential of stereoscopic observations for studying solar activities, the concept of "Solar Ring" mission containing six spacecraft at a sub-AU orbit in the ecliptic plane for panoramic solar observation was proposed\citep{Wang2020}. A variety of instruments are suggested onboard, including the white-light imager. One aim of the mission is to utilize multi-perspective white-light observations to track and reconstruct solar-wind transients. For this objective, how to deploy the six spacecraft or how to set the separation angle between two spacecraft to get the best reconstruction results of solar wind structures in the inner heliosphere is an important issue. In the current design, the six spacecraft are preliminarily divided into three groups in each of which the two spacecraft are separated by the angle of $30^{\circ}$, and the groups are separated by $120^{\circ}$ (see Figure \ref{fig:2-1})\citep{Wang2020b,Wang2020}. This scheme provides various separation angles of two spacecraft including $30^{\circ}$, $90^{\circ}$, $120^{\circ}$ and $150^{\circ}$, leading to various modes of multi-viewpoint observation. In this paper, we will study whether or not these angles are suitable for our CORAR method to reconstruct the solar wind structures and which one is the best.  
	
	\section{Methods}
	
	To look for the best satellite scheme for multi-viewpoint observation, the results of 3D reconstruction of artificial blob-like structures in synthetic dual images, obtained from dual spacecraft at a variety of separation angle, are investigated. The method is the same as that in \citet{Li2019}, and is described below.
	
	\subsection{Synthetic White-light Images}
	
\begin{figure*}[htb]
		\centering
		\includegraphics[height=0.36\columnwidth,width=0.9\columnwidth]{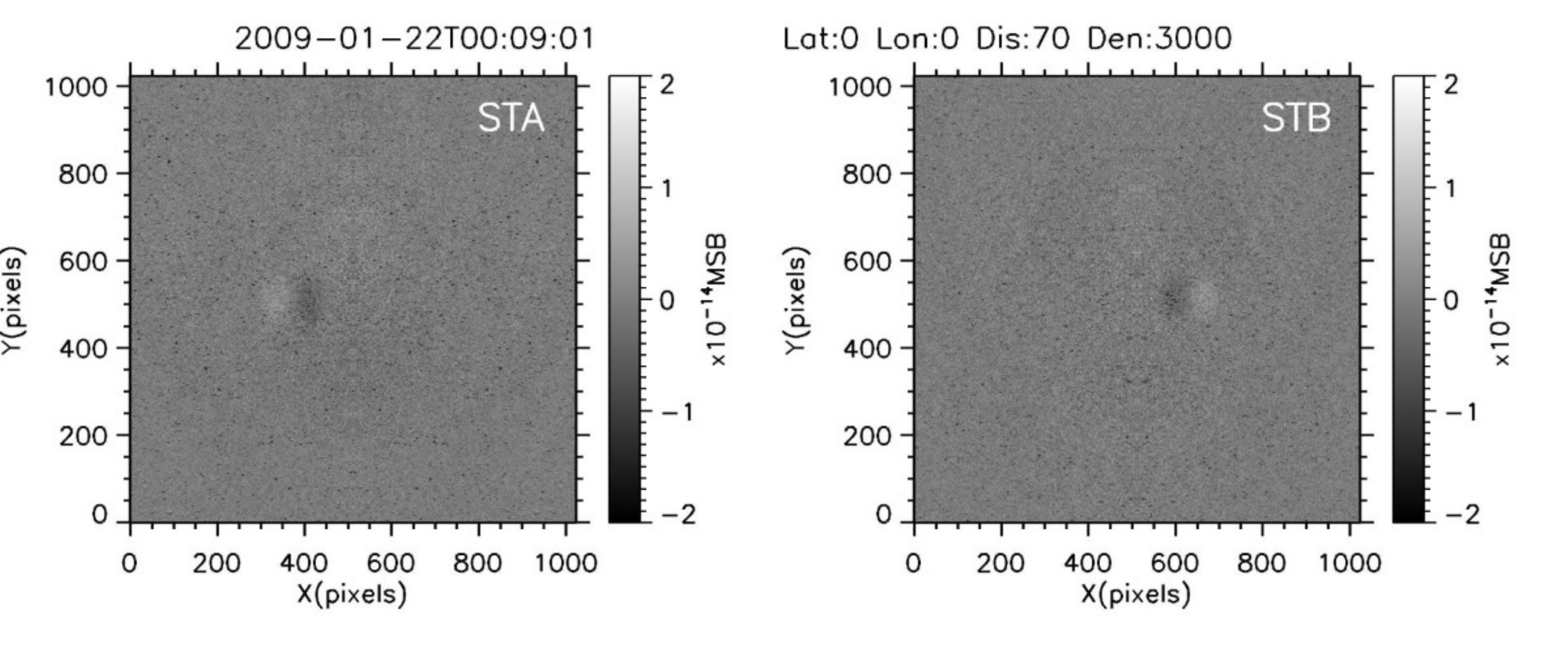}
		\caption{The synthetic HI-1 running-difference images containing the artificial blob from STEREO-A (left) and B (right) when the seperation angle is $90^{\circ}$. The location of the blob is $0^{\circ}$ in \textcolor[rgb]{0,0,0}{HEE} latitude, $0^{\circ}$ in \textcolor[rgb]{0,0,0}{HEE} longitude and 70$R_\odot$ in distance. The density of the blob center is $3\times 10^3 cm^{-3}$.}
		\label{fig:1}
	\end{figure*}

	
	As our 3D inversion method is based on dual-viewpoint white-light observations of solar wind, a large common \textcolor[rgb]{0,0,0}{space in the} FOV of two observers is required. The images from the Heliospheric Imager-1 (HI-1) onboard STEREO-A and B are chosen for our investigation. \textcolor[rgb]{0,0,0}{The two spacecraft orbits the Sun at an angular velocity of about $22^{\circ}/year$ faster or slower than the Earth}, providing a large FOV of $20^{\circ}\times 20^{\circ}$, capable of observing the space from $4^{\circ}$ to $24^{\circ}$ outward from the Sun. According to the plan of the Solar Ring mission, four cases of seperation angles, $60^{\circ}$, $90^{\circ}$, $120^{\circ}$ and $150^{\circ}$ are selected, of which the corresponding dates for STEREO are July 9 2008, January 22 2009, October 10 2009 and August 3 2010, respectively. The separation angle of $30^{\circ}$ is not \textcolor[rgb]{0,0,0}{discussed in this paper because of the small common region in FOV}. Figure \ref{fig:1} shows one pair of the synthetic white-light running-difference HI-1 images prepared for CORAR inversion method. In the images, a simulated blob-like structure \textcolor[rgb]{0,0,0}{based on Thomson Scattering theory} is inserted into an observational background image as a target of interest. The background image containing noise is generated by real images during solar quiescent periods without any notable eruptions. The simulated blob is set to have a cosine distribution\citep{Li2019} with a half-width of $R_0=5R_\odot$(solar radii) for all cases, with the maximum density of $1\times 10^3 cm^{-3}$, $3\times 10^3 cm^{-3}$, $1\times 10^4 cm^{-3}$ and $1\times 10^5 cm^{-3}$, respectively, at the blob center. The first two cases of density are closer to observational data, compared with the other two in the extreme or unreal condition \textcolor[rgb]{0,0,0}{which are studied for reducing the effect of low signal-noise ratio(SNR) and calculating correction curves (introduced in Section 2.2)}. For the separation angle of $120^{\circ}$ or $150^{\circ}$, we set a 3D mesh \textcolor[rgb]{0,0,0}{of blobs in the Heliocentric Earth Ecliptic(HEE) Coordinate} with the \textcolor[rgb]{0,0,0}{cell} size of $10R_\odot$ in radial direction, $10^{\circ}$ in both latitudinal and longitudinal directions. For the separation angle of $90^{\circ}$, the \textcolor[rgb]{0,0,0}{cell} size is changed to $5R_\odot \times 10^{\circ}\times 10^{\circ}$ since the \textcolor[rgb]{0,0,0}{volume of the }common FOV is much smaller than the cases of $120^{\circ}$ or $150^{\circ}$. For the separation angle of $60^{\circ}$, the \textcolor[rgb]{0,0,0}{cell} size is set smaller: $5R_\odot \times 5^{\circ}\times 5^{\circ}$. \textcolor[rgb]{0,0,0}{In one simulation, we reconstruct one blob in the mesh, and repeat the reconstruction procedure until all the grid cells have been gone through.}

	
	\subsection{3D reconstruction by CORAR method}
	
	The CORrelation-Aided Reconstruction (CORAR) method was developed for 3D reconstruction of inhomogeneous solar wind structures\citep{Li2018,Li2019}. It is assumed that if a non-uniform structure is located in the common FOV of observers from different perspectives, \textcolor[rgb]{0,0,0}{its} patterns in the images should be highly correlated \textcolor[rgb]{0,0,0}{when projected on the accurate position of the structure}. The reconstructed structure is presented using Pearson correlation coefficient ($cc_0$ in \textcolor[rgb]{0,0,0}{Eq.(\ref{eq:0})})
	
	\begin{equation}
	cc_0=\frac{\sum_{i=1}^{n}\left(x_i-\overline{x}\right)\left(y_i-\overline{y}\right)}{\sqrt{\sum_{i=1}^{n}\left(x_i-\overline{x}\right)^2}\sqrt{\sum_{i=1}^{n}\left(y_i-\overline{y}\right)^2}}
	\label{eq:0}
	\end{equation}   
	
	\noindent where $x_i$ and $y_i$ represent \textcolor[rgb]{0,0,0}{the data of projected images from two perspectives respectively} within the sampling area, a box running over the entire common FOV \textcolor[rgb]{0,0,0}{in polar coordinate}, and $\overline{x}$,$\overline{y}$ are the related averages. Before the calculation of the \textcolor[rgb]{0,0,0}{$cc_0$} value, the HI-1 images from STEREO-A and B are projected onto the same meridian plane. In this procedure, the \textcolor[rgb]{0,0,0}{HEE coordinate} is used, and the grid \textcolor[rgb]{0,0,0}{element} is set to $1^{\circ}$ in latitudinal direction and about $0.08R_\odot$ in radial direction. The size of the sampling area is 41 (about $3.28R_\odot$ in distance)$\times$11 (about $11^{\circ}$ in latitude). The meridian plane, on which the HI-1 images are projected, is scanned along the longitude by $1^{\circ}$ to get the 3D distribution of the \textcolor[rgb]{0,0,0}{$cc_0$} value.
	

	In the images, some relatively dim structures such as those at a larger distance may have a lower \textcolor[rgb]{0,0,0}{SNR}. Those features are relatively inconspicuous and lead to low \textcolor[rgb]{0,0,0}{$cc_0$} values during the process of correlation calculation. As the consequence, they may be judged as non-real structures. To solve this problem, the \textcolor[rgb]{0,0,0}{$cc_0$} values are further corrected in accordance with the relationship between \textcolor[rgb]{0,0,0}{$cc_0$} and total signal-to-noise ratio (TSNR), a parameter determined by SNR from dual images. TSNR and SNR are calculated as follows
	
	\begin{align}
	TSNR & =\sqrt{SNR_a^2+SNR_b^2}\\
	SNR_{a,b} & =\sqrt{\frac{\sum \frac{\left(X_{a,b}-\overline{X_{a,b}}\right)^2}{N}}{\sum \frac{\left(X_{a,b}^0-\overline{X_{a,b}^0}\right)^2}{N}}-1}
	\label{eq:3}
	\end{align}

\begin{figure*}[htbp]
		\centering
		\includegraphics[height=11cm,width=16cm]{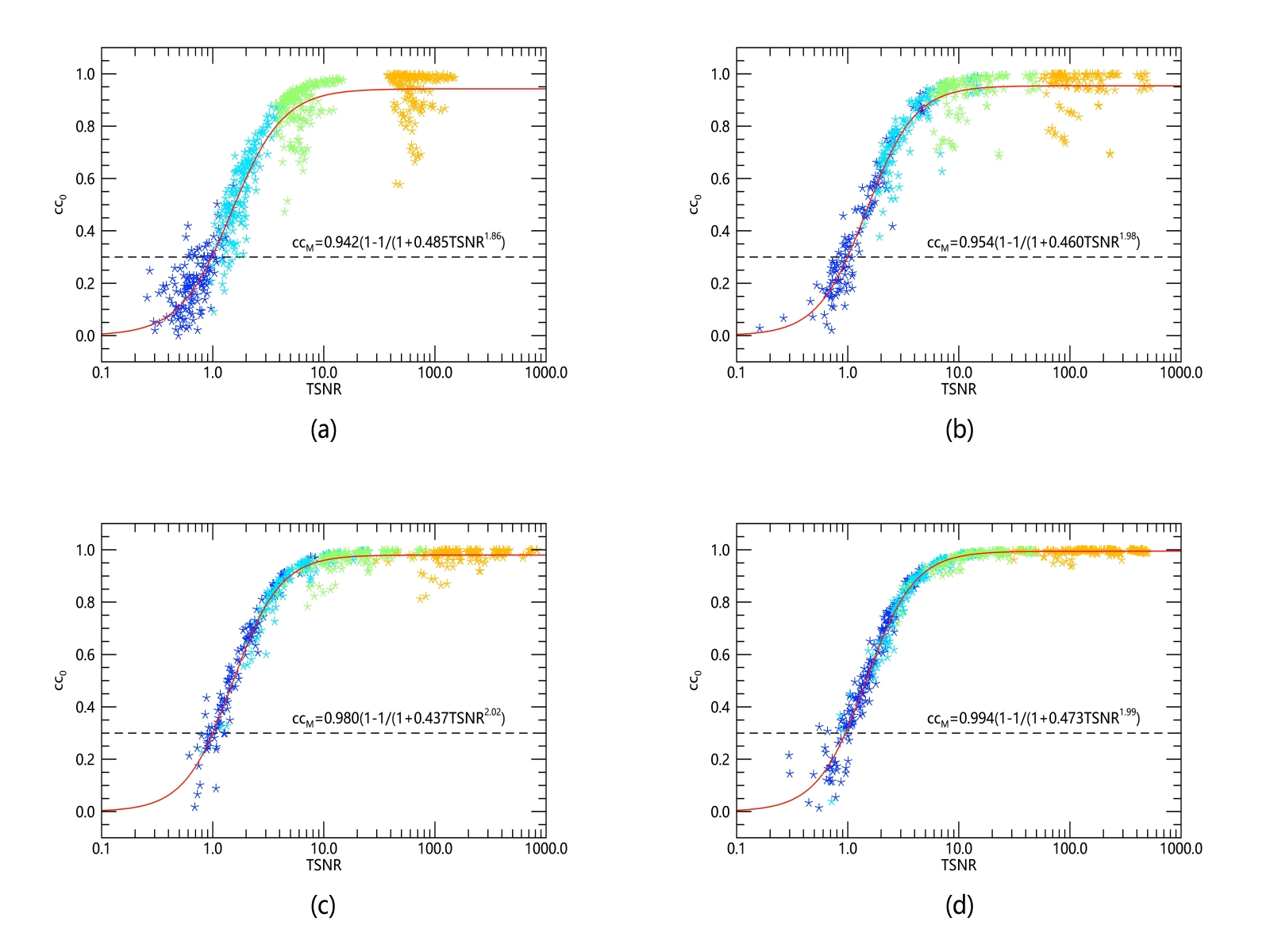}
		\caption{The correction curve for cc calculation process. The panels from (a) to (d) are respectively from cases of $60^{\circ}$, $90^{\circ}$, $120^{\circ}$ and $150^{\circ}$. The red curves represent the fitting profiles with their functions listed in each panel, and the threshold 0.3 is marked using black dotted lines. Data from blobs with the density of $1\times 10^3 cm^{-3}$,  $3\times 10^3 cm^{-3}$,  $1\times 10^4 cm^{-3}$ and  $1\times 10^5 cm^{-3}$ is relatively labelled with dark blue, sky blue, green and orange.}
		\label{fig:2}	
	\end{figure*}

	\noindent $X_{a,b}$ and $X_{a,b}^0$ represent data of the artificial blob and background noise, respectively, in the sampling area (see the paper\citep{Li2019} for details). The relationship between \textcolor[rgb]{0,0,0}{$cc_0$} and TSNR at the center of blobs for the four cases with different separation angles are shown in Figure \ref{fig:2}, in which different colors represent different density of blobs. The red curves in these figures, called the correction curves, are the fitting curves of data using the following formula
	
	\begin{align}
	cc_M=A\left(1-\frac{1}{1+B \cdot TSNR^C}\right)
	\label{eq:5}
	\end{align}
	
	\noindent $A$, $B$ and $C$ are the fitting parameters and their initial values are $1$, $0.5$ and $2$, respectively. The values of these fitting parameters are marked on Figure \ref{fig:2}. The correction profiles with fitting values close to the initial parameters for four cases indicate that the relationship between \textcolor[rgb]{0,0,0}{$cc_0$} and TSNR does not vary significantly with the separation angle. But a trend is obvious that the scattered points are more concentrated around the red line for wider separation angle. To correct the calculated \textcolor[rgb]{0,0,0}{$cc_0$} value, we compare the initial value $cc_0$ with the the threshold (marked with black dotted lines in Figure \ref{fig:2}), which nearly represents TSNR=1 in the correction curves of four cases. If $cc_0$ is above the threshold, we calculate the corresponding TSNR according to the sampling area and get $cc_M$ from the correction curve. The corrected \textcolor[rgb]{0,0,0}{value $cc$} is just $cc_0/cc_M$. 
	
	
	In the case of $60^{\circ}$, the overall distribution of \textcolor[rgb]{0,0,0}{$cc_0$} values is lower than in other cases, especially for blobs with the lowest density $10^3 cm^{-3}$ (dark blue points in Figure \ref{fig:2}), of which most have \textcolor[rgb]{0,0,0}{$cc_0$} values lower than the threshold. \textcolor[rgb]{0,0,0}{It would be difficult to reconstruct these blobs even after the cc correction process}. One reason for this phenomenon is that the angle between the two spacecraft is small, and blobs in their common FOV \textcolor[rgb]{0,0,0}{possibly} propagate near the direction toward one of the spacecraft. In this case, the blob signature will be subtracted in the running-difference images, resulting an extremely low cc value. The \textcolor[rgb]{0,0,0}{second} reason is the angle between the LOS of the two HI-1 cameras is close to $90^{\circ}$. \textcolor[rgb]{0,0,0}{Thus, the Thomson surfaces\citep{Howard2012,DeForest2013,Howard2013} of the two observers are nearly perpendicular at the intersection.} In this case, most blobs will locate far from one of the Thomson surfaces, and may be poorly reconstructed because of their faint features in the images. Therefore, we suspect that the scheme with a small seperation angle, say about $60^{\circ}$, may not be suitable for our 3D reconstruction method. This will be further verified in section 3.

\begin{figure*}[htb]
		\centering
		\subfigure[]{
			\includegraphics[height=0.5\columnwidth,width=0.8\columnwidth]{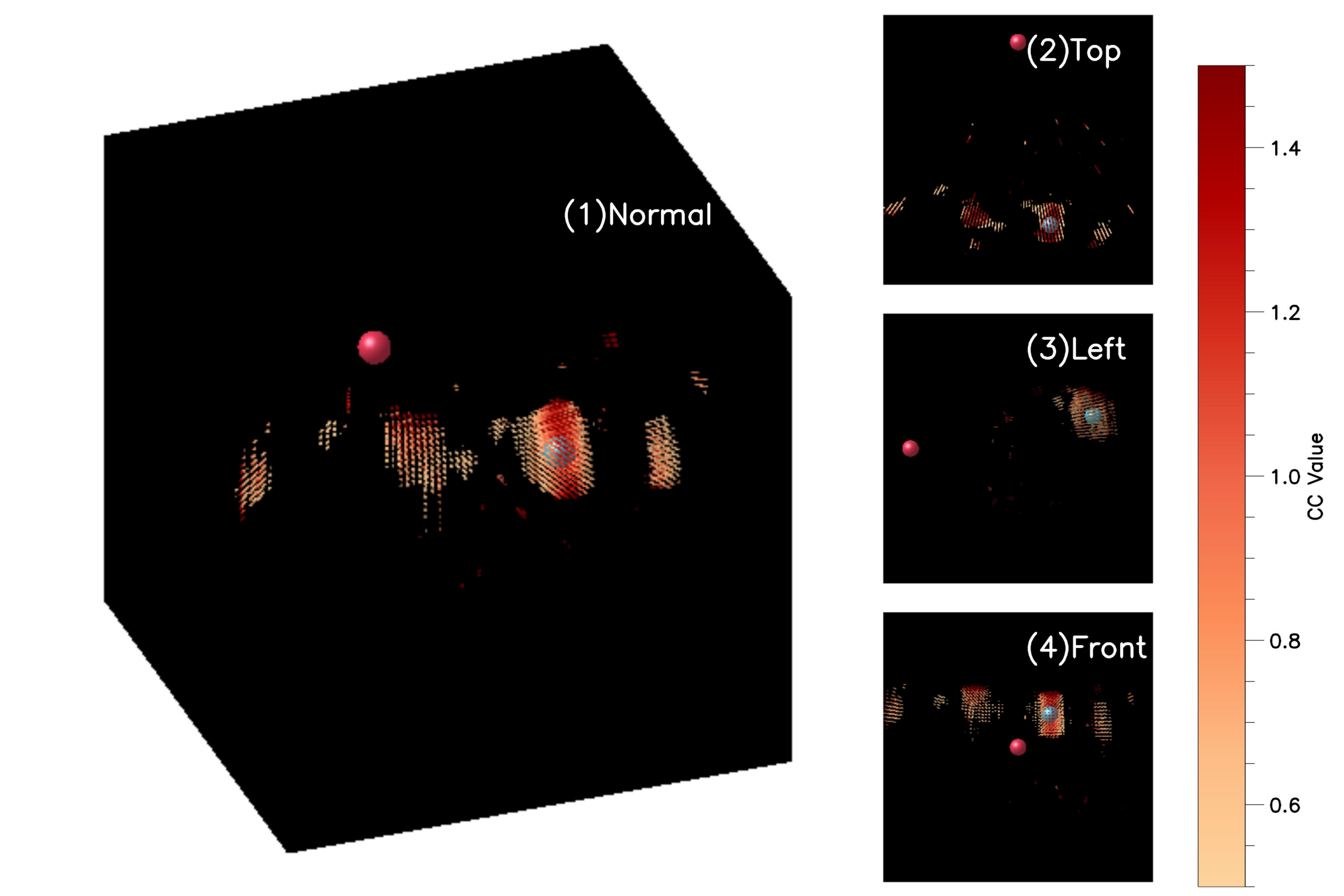}
		}
		\subfigure[]{
			\includegraphics[height=0.5\columnwidth,width=0.8\columnwidth]{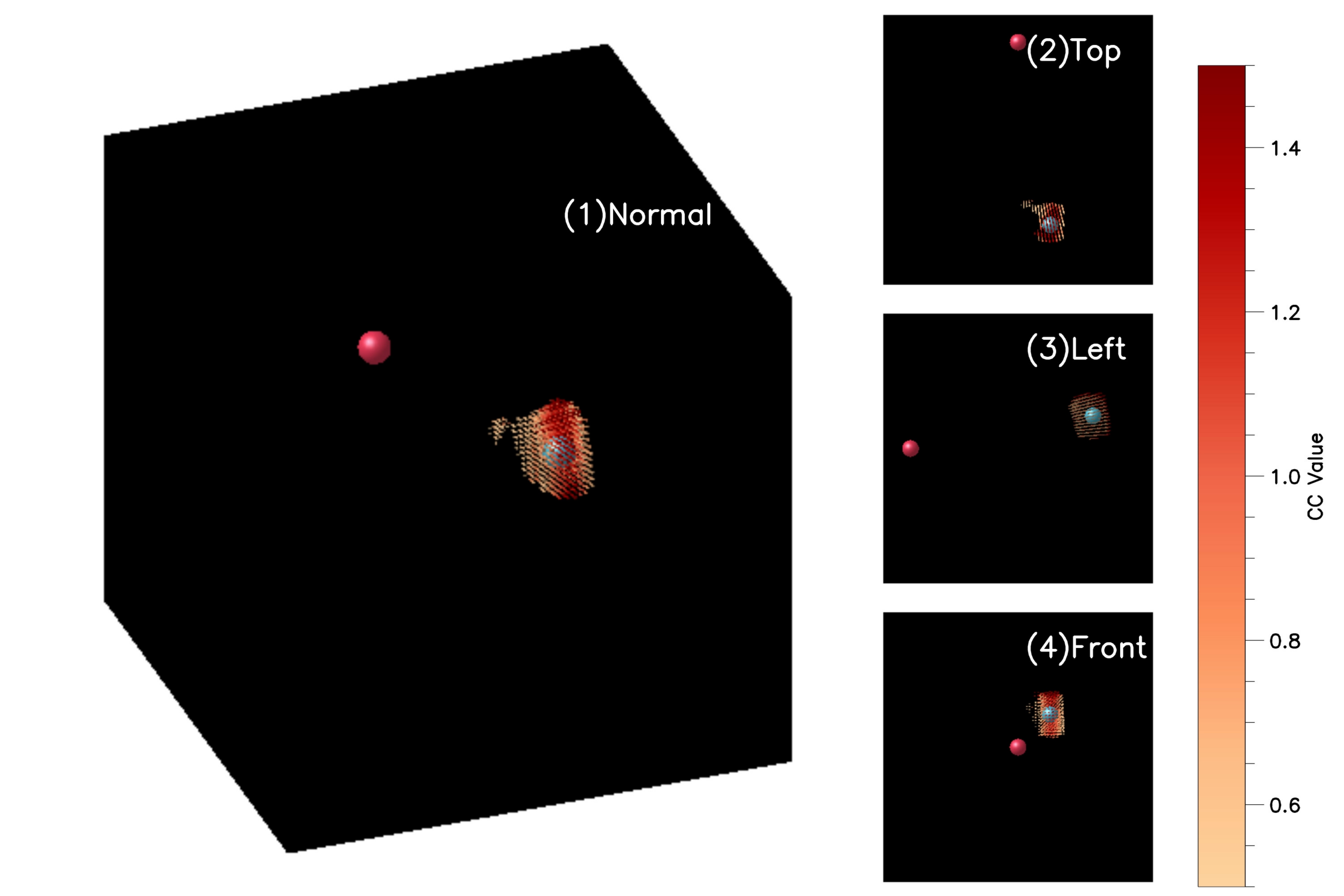}
		}	
		\caption{Panel (a): The 3D cc map showing the reconstructed high-cc regions from different perspectives, with the green ball marking \textcolor[rgb]{0,0,0}{the location of} the simulated blob and the red ball for the Sun. Panel (b): The same as Panel (a) with all the small regions removed. In this example, the seperation angle of the two spacecraft is $120^{\circ}$, the location of the simulated blob in HEE is $10^{\circ}W ,10^{\circ}N$ and 70$R_\odot$, and the central density is $3000cm^{-3}$.}
		\label{fig:3}
	\end{figure*}

	
	All the regions with the corrected cc value larger than or equal to 0.5, called high-cc regions, are considered \textcolor[rgb]{0,0,0}{as} recognized features. Although the above cc correction enhances the capability in recognizing faint features , it may result in some small "fake" structures. In our study here, the largest high-cc region is further selected as the reconstructed blob (shown in Figure \ref{fig:3}).  
	
	\subsection{\textcolor[rgb]{0,0,0}{Method assessment}}
	
	
	To evaluate the performance of the reconstruction of simulated blobs, we define two parameters: the Positional Deviation (PD) and the Expansion Ratio (ER). PD is the offset distance of the reconstructed structure from the simulated blob in one dimension, calculated by Eq.(\ref{eq:1})
	
	\begin{equation}
	\bigtriangleup x_i=\frac{\sum_{cc>0.5}cc\cdot \left(x_i-x_{i0}\right)}{\sum_{cc>0.5}cc},i=r,\theta,\varphi
	\label{eq:1}
	\end{equation}
	
	\noindent where $x_i,i=r,\theta,\varphi$ represent positional parameters in the 3D coordinate and $x_{i0}$ represent the accurate position of the blob center. ER is the ratio of the cc-weighted width of the reconstructed structure over that of the simulated blob, calculated by Eq.(\ref{eq:2})
	
	\begin{equation}
	\kappa_i=\sqrt{\frac{\sum_{cc>0.5}cc\cdot \left(x_i-x_{i0}\right)^2}{\sum_{cc>0.5}cc\cdot R_0^2}},i=r,\theta,\varphi
	\label{eq:2}
	\end{equation}
	
	\noindent ER measures the integrity of the inversion results and the possible expansion caused by fake structures or other reasons in each dimension. $x_i$ in the directions of latitude and longitude has the unit of $R_\odot$, and the angular positional parameters $x_{\theta}$ and $x_{\phi}$ are the angular components $\theta$ and $\phi$ multiplied by $R$ and $R cos\theta$, respectively.

	Considering that the radius of blobs is 5$R_\odot$, we use the criteria below to determine the goodness of the reconstruction

	\begin{equation}
	\begin{aligned}
	|\bigtriangleup x_i| & < 3R_\odot ,i=r,\theta,\varphi\\
	\kappa_i & <1.2,i=r,\theta,\varphi
	\label{eq:4}
	\end{aligned}
	\end{equation}
	
	The reconstructed blobs are divided into four performance levels:  
	\begin{itemize}
		\item Level 1: Reconstructed blobs completely satisfy Eq.(\ref{eq:4}) in 3 dimensions for all densities.
		\item Level 2: Reconstructed blobs do not fully \textcolor[rgb]{0,0,0}{satisfy} Eq.(\ref{eq:4}) when the density is as low as $1\times 10^3 cm^{-3}$.  
		\item Level 3: Reconstructed blobs only \textcolor[rgb]{0,0,0}{satisfy} the criteria for PD with a moderate or high density ($\ge 3\times 10^3 cm^{-3}$). 
		\item Level 4: It is the worst case that do not satisfy any of the above criteria. 
	\end{itemize}
	
	Based on the statistics of four performance levels of the reconstruction, the regions suitable for CORAR reconstruction with various spacecraft angles can be given.


	\section{Results}

\begin{figure*}[htb]
		\centering
		\includegraphics[height=0.6\columnwidth,width=0.86\columnwidth]{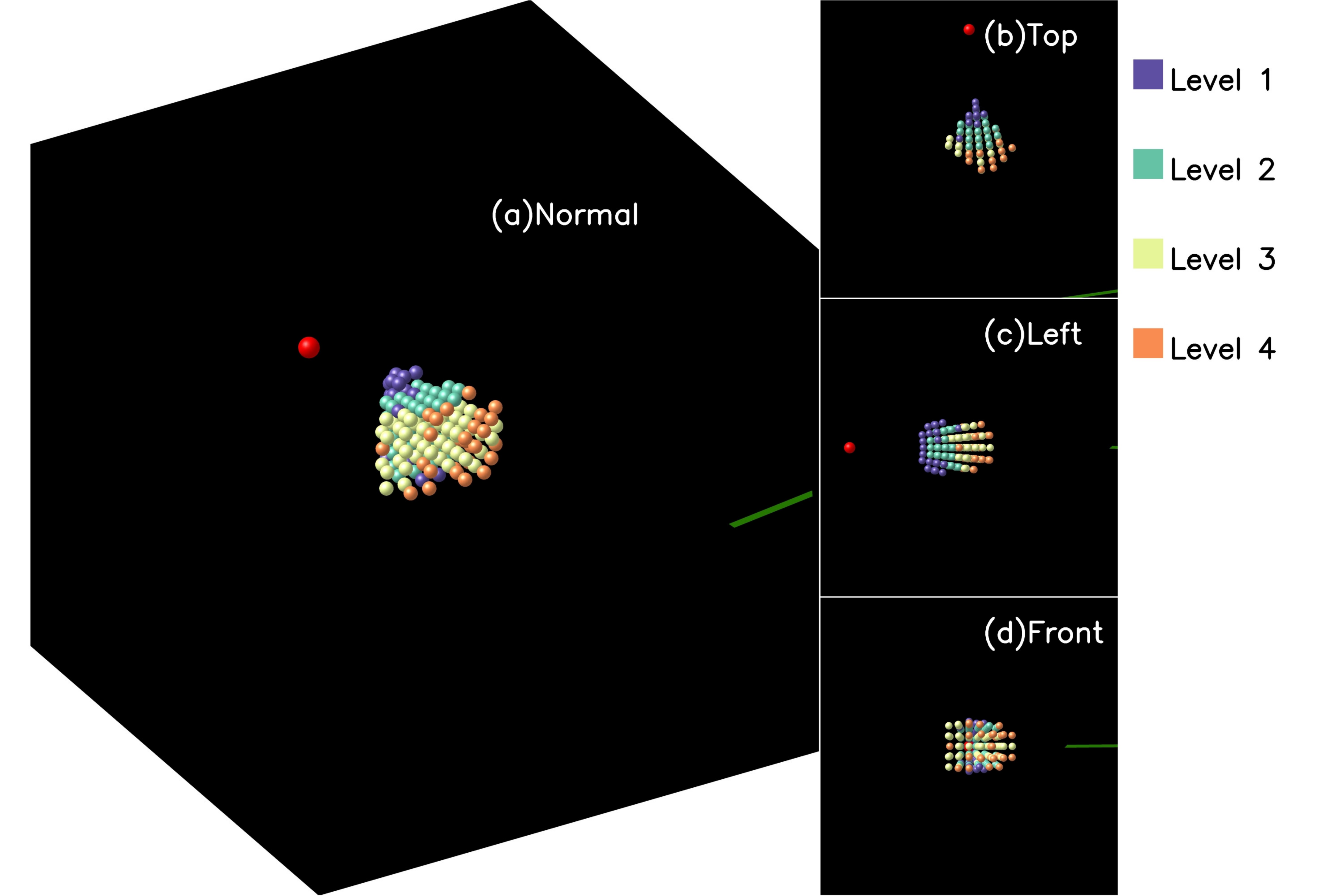}
		\caption{The 3D distribution of the reconstruction performance for the case that separation angle of the two spacecraft is $60^{\circ}$. The four levels, from Level 1 to 4, of the reconstruction performance are respectively marked with durk blue, green, yellow and orange. The Sun is the red ball and the green line represents the connection line of the two spacecraft.} 
		\label{fig:4-1}
	\end{figure*}

\begin{figure*}[htb]
		\centering
		\includegraphics[height=0.6\columnwidth,width=0.86\columnwidth]{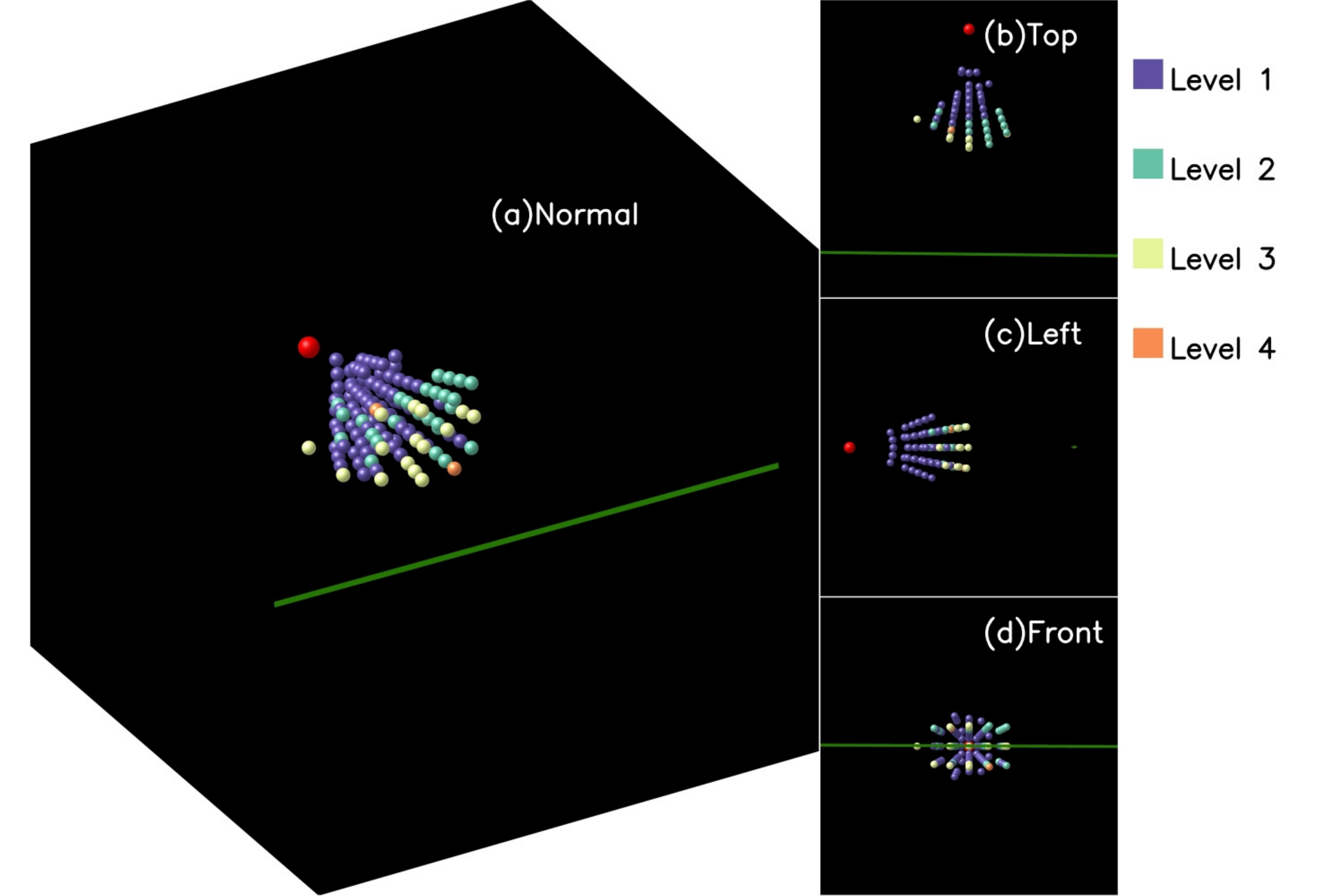}
		\caption{The 3D distribution of the reconstruction performance for the case of $90^{\circ}$. }
		\label{fig:4-3}
	\end{figure*}

\begin{figure*}[htb]
		\centering
		\includegraphics[height=0.6\columnwidth,width=0.86\columnwidth]{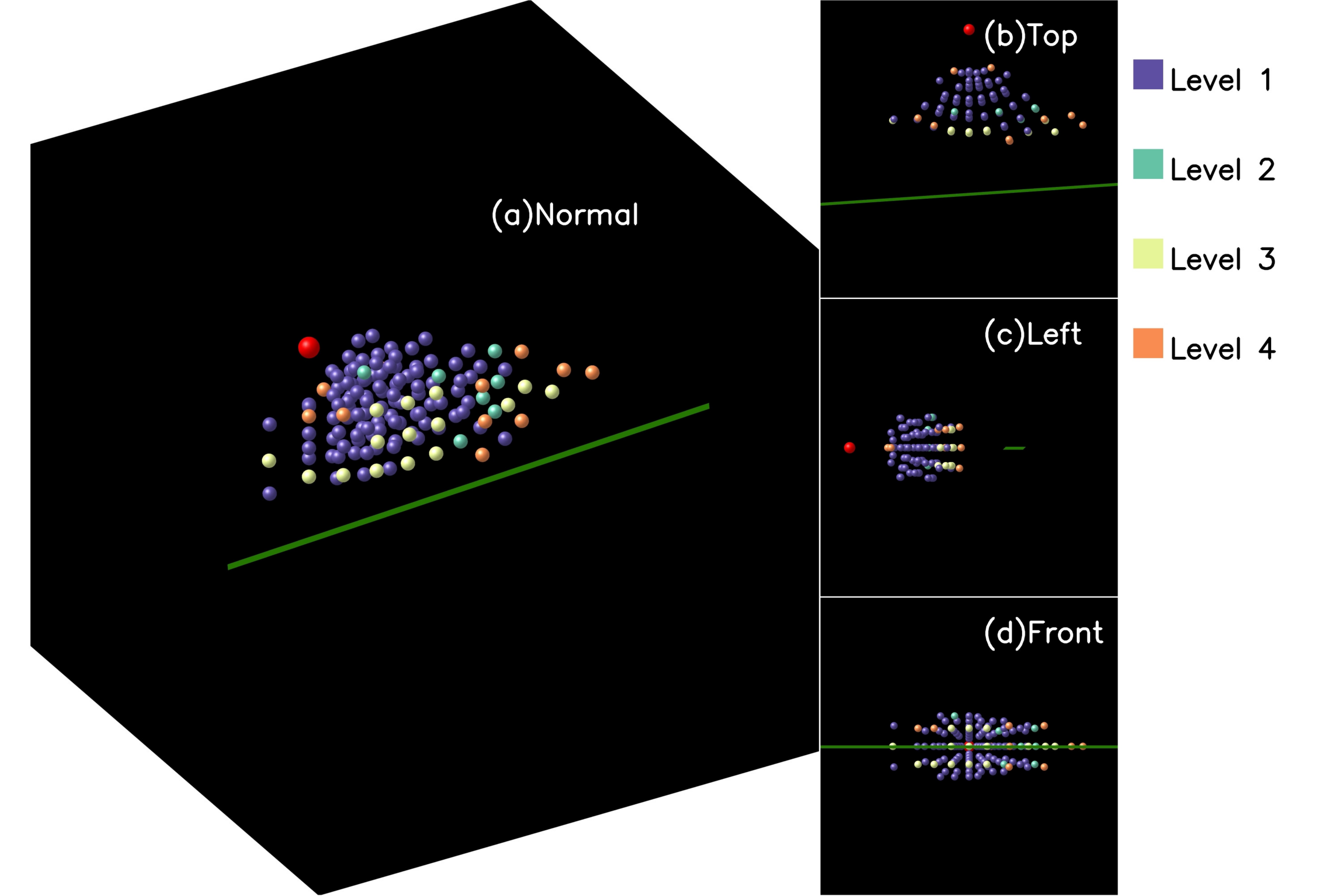}
		\caption{The 3D distribution of the reconstruction performance for the case of $120^{\circ}$.}
		\label{fig:4-5}
	\end{figure*}

\begin{figure*}[htb]
		\centering
		\includegraphics[height=0.6\columnwidth,width=0.86\columnwidth]{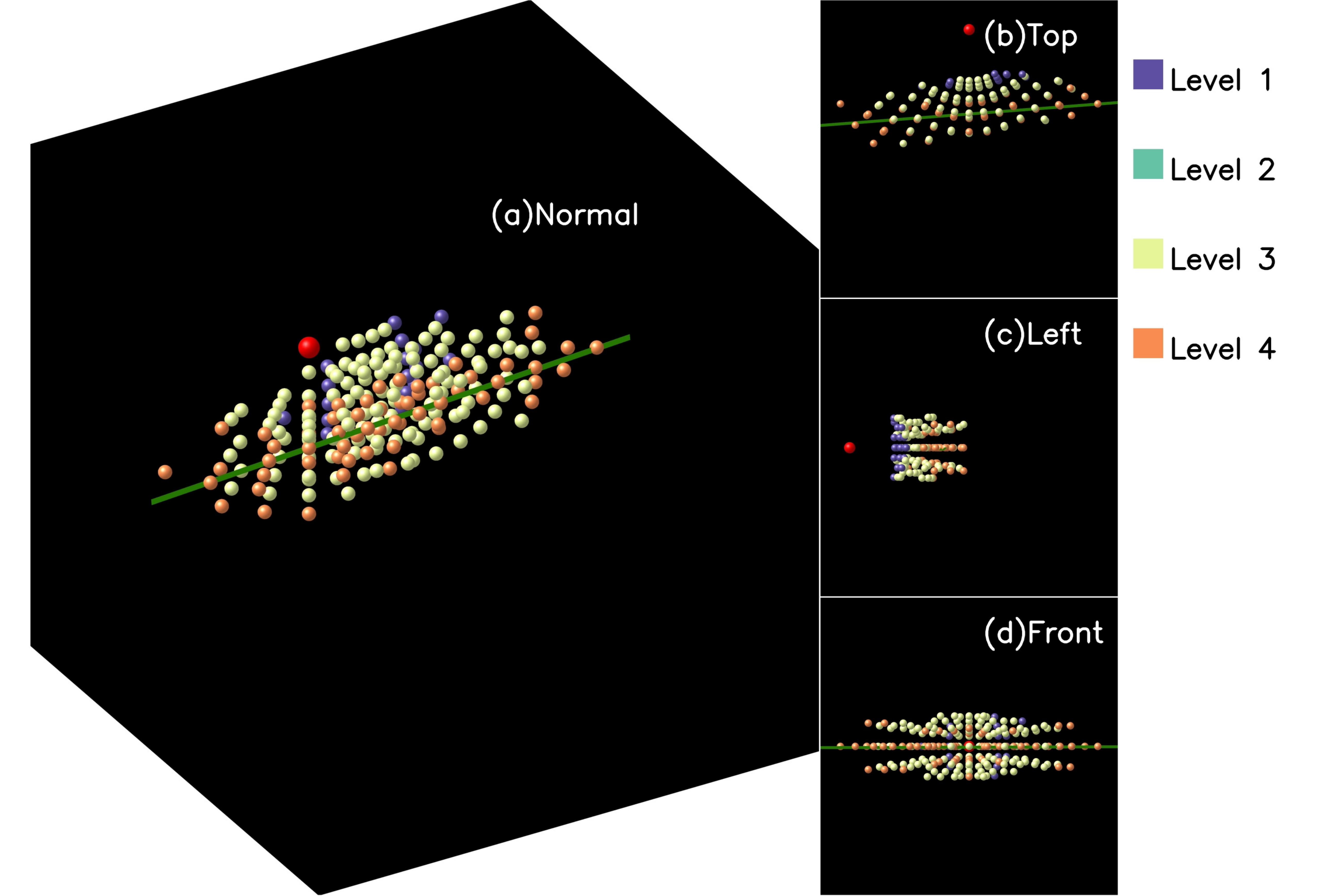}
		\caption{The 3D distribution of the reconstruction performance for the case of $150^{\circ}$.}
		\label{fig:4-7}
	\end{figure*}

\begin{figure*}[htb]
		\centering
		\subfigure[]{
			\includegraphics[height=0.3\columnwidth,width=0.4\columnwidth]{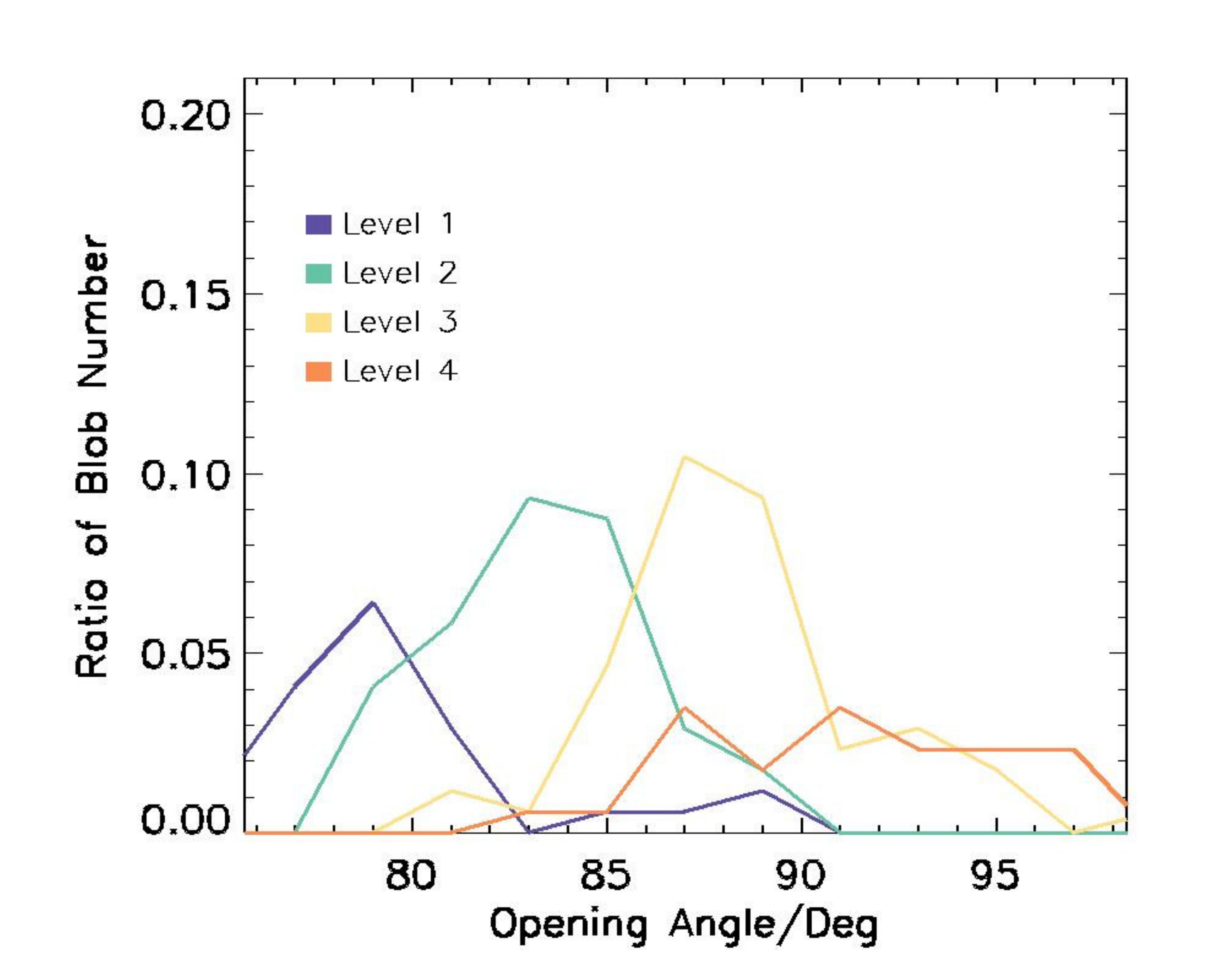}
		}
		\subfigure[]{
			\includegraphics[height=0.3\columnwidth,width=0.4\columnwidth]{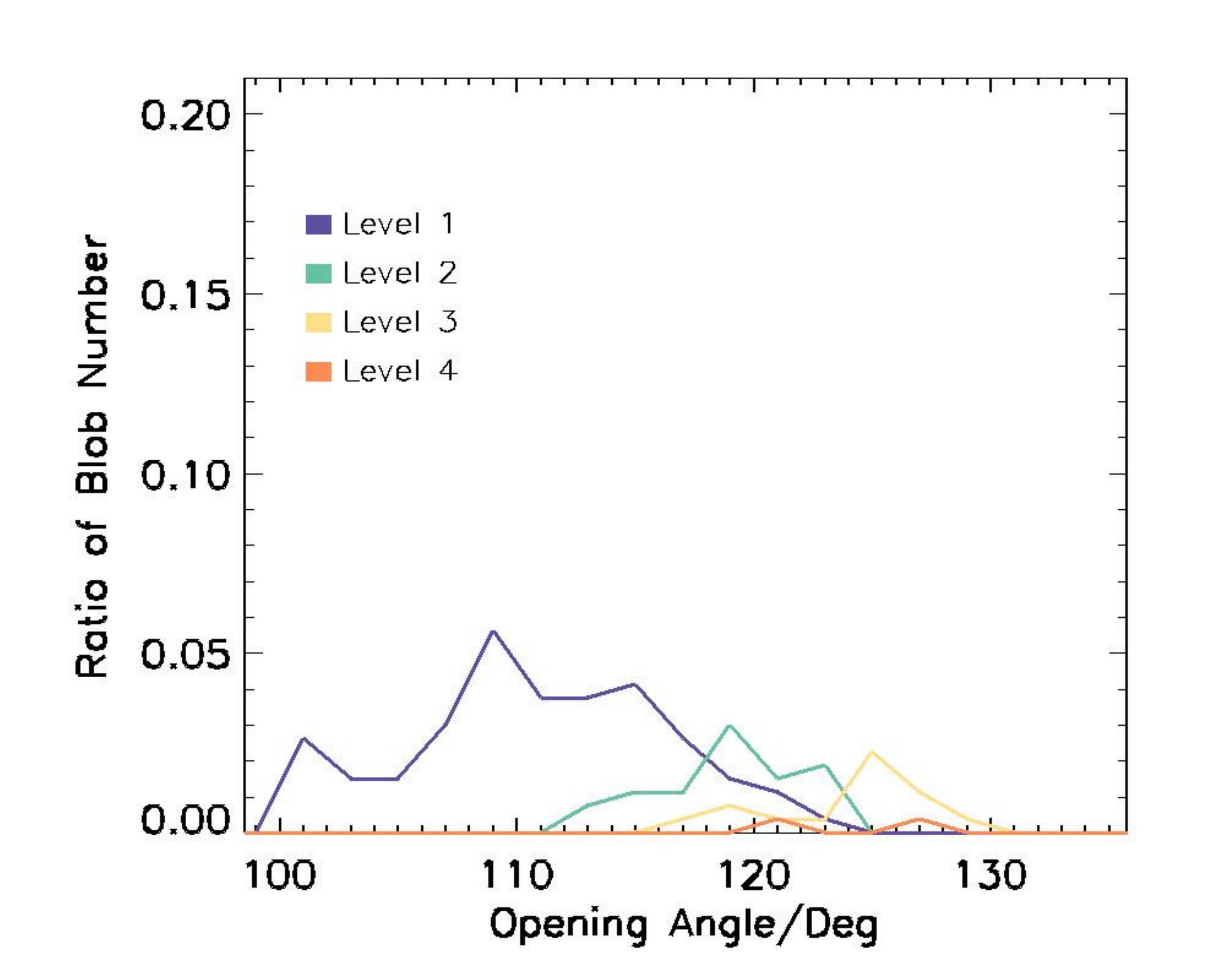}
		}
		\subfigure[]{
			\includegraphics[height=0.3\columnwidth,width=0.4\columnwidth]{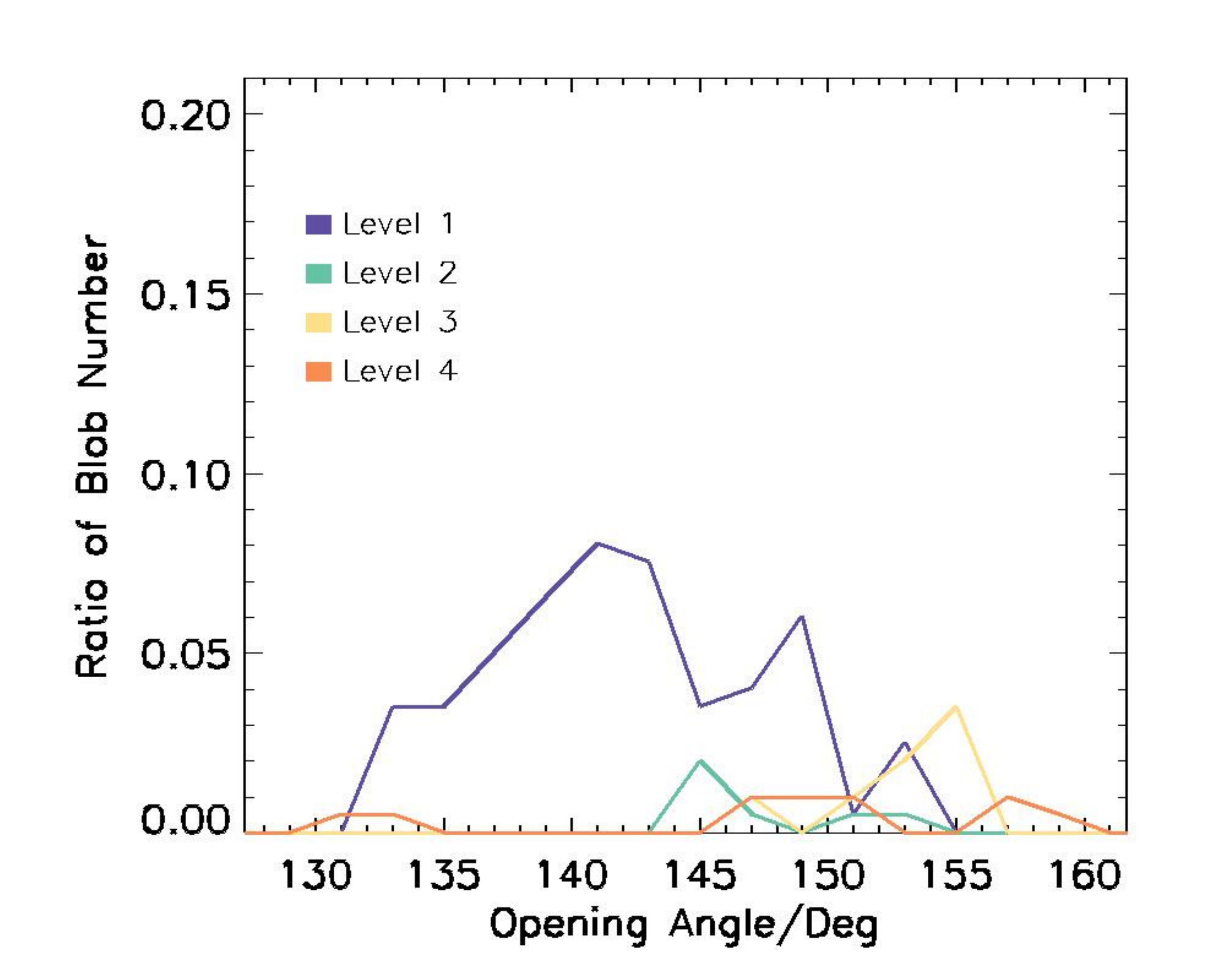}
		}
		\subfigure[]{
			\includegraphics[height=0.3\columnwidth,width=0.4\columnwidth]{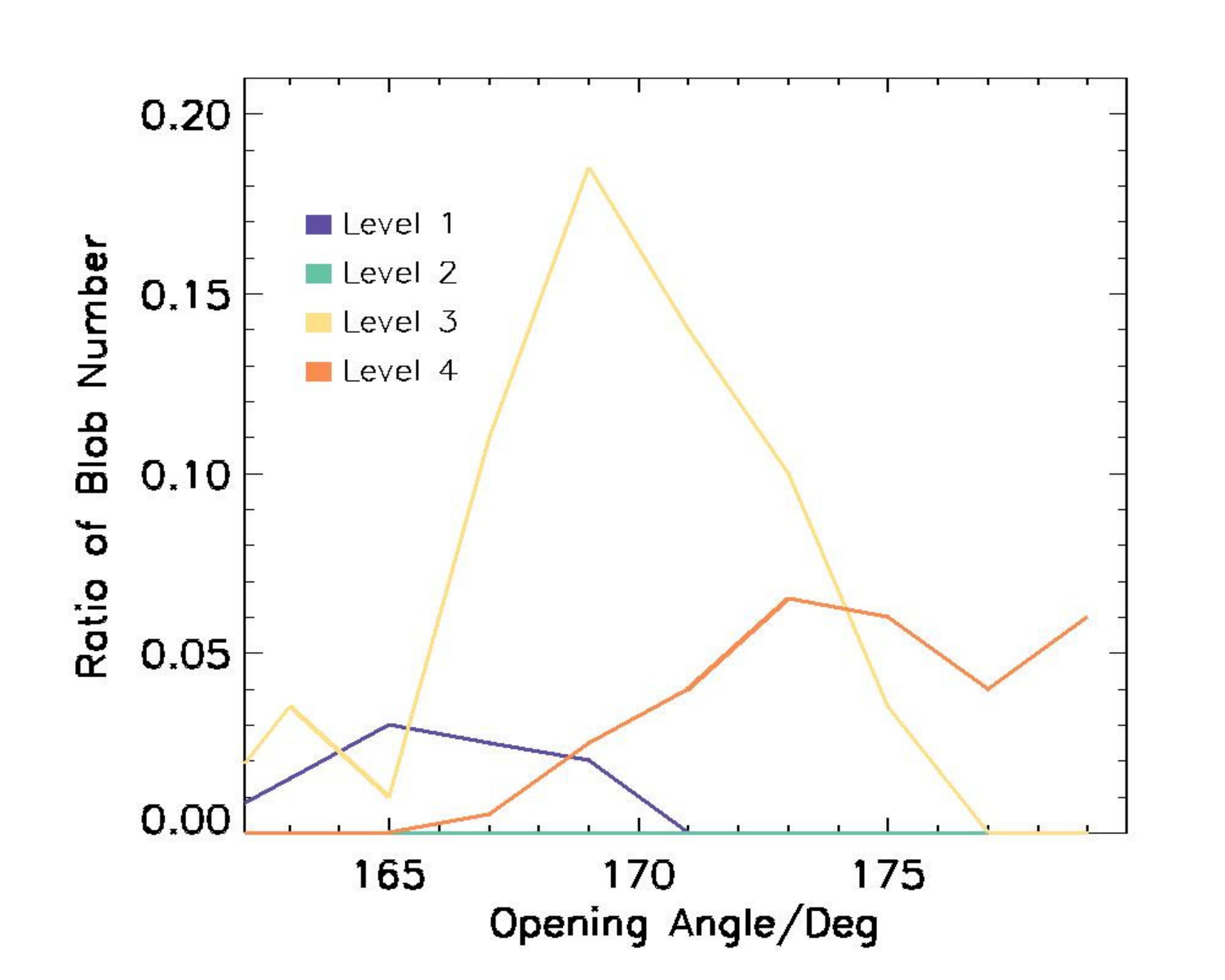}
		}
		\caption{The normalized distribution of the reconstruction performance for the case that separation angle of the two spacecraft is (a)$60^{\circ}$, (b)$90^{\circ}$, (c)$120^{\circ}$ and (d)$150^{\circ}$. The horizonal axis is the opening angle of the center of blobs to the two spacecraft, and the vertical axis is the ratio of the number of blobs with the certain angle over the number of all blobs.} 
		\label{fig:5}
	\end{figure*}

	
	With different stereoscopic angles, the total area of space covered by dual spacecraft varies. Generally, \textcolor[rgb]{0,0,0}{the volume of} the common FOV increases as the separation angle increases. This is beneficial for the reconstruction of solar transients, but also brings about negative effects, for example, the collinear effect\citep{Li2018}. For the Solar Ring mission, the common space with four different angles (i.e. $60^{\circ}$, $90^{\circ}$, $120^{\circ}$ and $150^{\circ}$) of dual spacecrafts are studied about their suitability for reconstruction respectively. Figure \ref{fig:4-1}-\ref{fig:4-7} demonstrate the distribution of the goodness of the reconstruction in the 3D space, in which four different colors represent the four levels of performance of the reconstruction. For further understanding the effect from the opening angle of blobs\textcolor[rgb]{0,0,0}{(the angle between the two lines connecting one blob and the two spacecraft respectively)} on the reconstruction, the angular distribution profiles of the reconstruction results are shown in Figure \ref{fig:5}, which are normalized by the total number of \textcolor[rgb]{0,0,0}{blobs under each separation-angle condition}. This factor also measures the distance from blobs to the connecting line of spacecraft, related to the collinear effect in cases of larger separation angle. \textcolor[rgb]{0,0,0}{The spatial parameters like the ranges of common FOV and opening angle applied in our work are summarized in Table \ref{tab1}.}
	
	\begin{table}[htb]
		\centering\small
		\caption{The spatial parameter range under four angular conditions}
		\label{tab1}
		\begin{tabular}{lllll}
			\toprule
			Separation angle   & $60^{\circ}$  & $90^{\circ}$ & $120^{\circ}$ & $150^{\circ}$    \\
			\midrule
			Distance($R_\odot$) & (30.9,101.4) & (19.8,92.8) & (17.8,85.4) & (15.9,87.1) \\
			Latitude(deg) & (-45,45) & (-57,57) & (-63,63) & (-66,66) \\
			Longitude(deg) & (-18,27) & (-39,36) & (-54,60) & (-66,74) \\
			Opening angle(deg) & (75,98) & (98,136) & (127,162) & (162,180) \\
			\bottomrule
		\end{tabular}
	\end{table}

	At a separation angle of $60^{\circ}$, the total volume of the common FOV in our study is the smallest among four cases, with the distance between about $30.9R_\odot$ and $101.4R_\odot$ along the Sun-Earth line, latitude between about $\pm 45^{\circ}$, and longitude between $-18^{\circ}$ and $27^{\circ}$ in our investigation. Besides, in this case, the angle between the LOS from dual spacecrafts to the observed space is near $90^{\circ}$, meaning the smallest common region near the Thomson surfaces of the two spacecraft as mentioned before. As shown in Figure \ref{fig:4-1}, the best reconstructed blobs are concentrated in the region at the smallest distance, which are marked \textcolor[rgb]{0,0,0}{in purple}. As the distance from the Sun increases, there \textcolor[rgb]{0,0,0}{is an} decreasing trend of the performance level. Furthermore, it can be found in Figure \ref{fig:5}(a) that the opening angles of most blobs with respect to the two spacecraft are about between $75^{\circ}-100^{\circ}$, far away from the connecting line of two observers, and blobs with a smaller angle have better inversion results. It is noted that at large opening angle, especially $>85^{\circ}$, the reconstruction has the lowest level of performance.

	%
	%
	

	When the angle of dual spacecrafts is $90^{\circ}$, the common FOV is between $19.8R_\odot$ and $92.8R_\odot$ in distance along the Sun-Earth line, $-57^{\circ}$ and $57^{\circ}$ in latitude, and $-39^{\circ}$ and $36^{\circ}$ in longitude. The area of available space for reconstruction is larger than that of $60^{\circ}$ case. The 3D distribution of the reconstruction performance is shown in Figure \ref{fig:4-3}. Unlike the case of $60^{\circ}$, most reconstructed blobs are at the best performance level. \textcolor[rgb]{0,0,0}{On the other hand, there is a decreasing trend} of performance level in the far-distance region \textcolor[rgb]{0,0,0}{similar to the $60^{\circ}$ case}. The map of angular distribution in Figure \ref{fig:5}(b) shows the opening angle of blob ranges from $100^{\circ}$ to $130^{\circ}$, with a small part of poorly reconstructed blobs distributing above $120^{\circ}$. 

	The case of $120^{\circ}$ angle has a larger space available for reconstruction. The common FOV is from $17.8R_\odot$ to $85.4R_\odot$ in distance along the Sun-Earth line, $-63^{\circ}$ to $63^{\circ}$ in latitude, and $-54^{\circ}$ to $60^{\circ}$ in longitude. As the separation angle gets larger, the common FOV in the outer space is closer to the connection line of the two spacecraft. It leads to a severe phenomenon called "collinear effect"\citep{Li2018} that the structure near the line connecting the two observers is difficult to be precisely located. This problem is due to the inherent drawbacks of double-viewpoint observations and may cause the expansion of structures along the connecting line. Similar to the case of $90^{\circ}$, the main part of the common FOV belongs to the best performance level (see Figure \ref{fig:4-5}), and the outer edge belongs to the worst level. As shown in Figure \ref{fig:5}(c), the maximum opening angle of blobs to the spacecraft is about $160^{\circ}$, which is nearly parallel, and most reconstructed blobs with opening angle above $150^{\circ}$ are not good enough.

	In the case of the largest angle in our study, the LOS of dual spacecrafts is almost parallel and the connection line appears in the FOV of HI-1 instrument. The common space used for the reconstruction is the largest, with distance ranging from $15.9R_\odot$ to $87.1R_\odot$ along the Sun-Earth line, latitude from $-66^{\circ}$ to $66^{\circ}$, longitude from $-66^{\circ}$ to $74^{\circ}$. In the case of $150^{\circ}$, most of the common space is not suitable for the reconstruction, especially in the vicinity of the connection line where the blobs at Level 3 or Level 4 are widely distributed (see Figure \ref{fig:4-7}). In the pattern of the angular distribution, as revealed in Figure \ref{fig:5}(d), only a few blobs with the opening angle smaller than $170^{\circ}$ at small distance are well reconstructed, whereas other blobs are poorly reconstructed. The collinear effect seriously influences the reconstruction performance.

\begin{figure*}[htb]
		\centering
		\subfigure[]{
			\includegraphics[height=0.3\columnwidth,width=0.4\columnwidth]{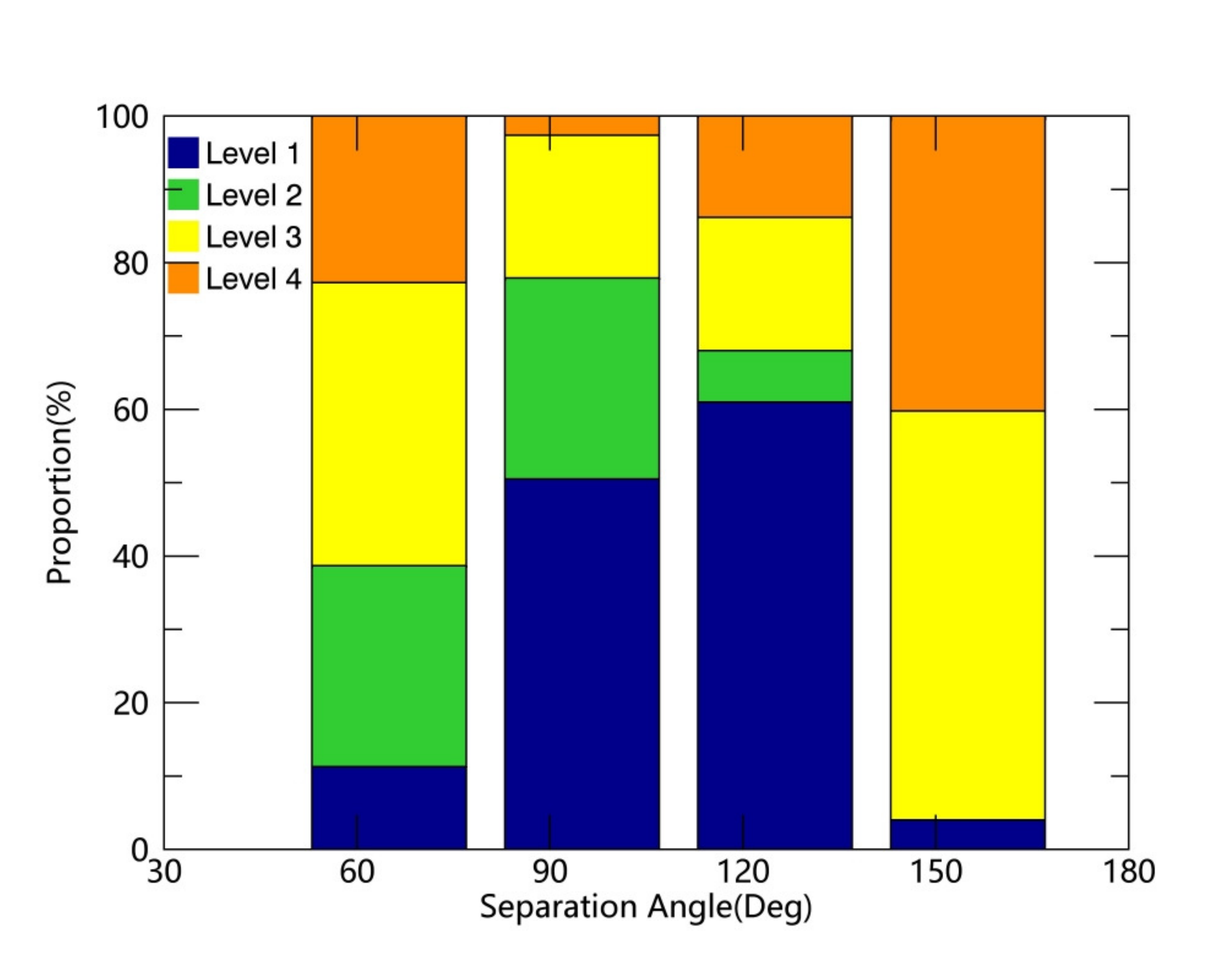}
		}
		\subfigure[]{
			\includegraphics[height=0.3\columnwidth,width=0.4\columnwidth]{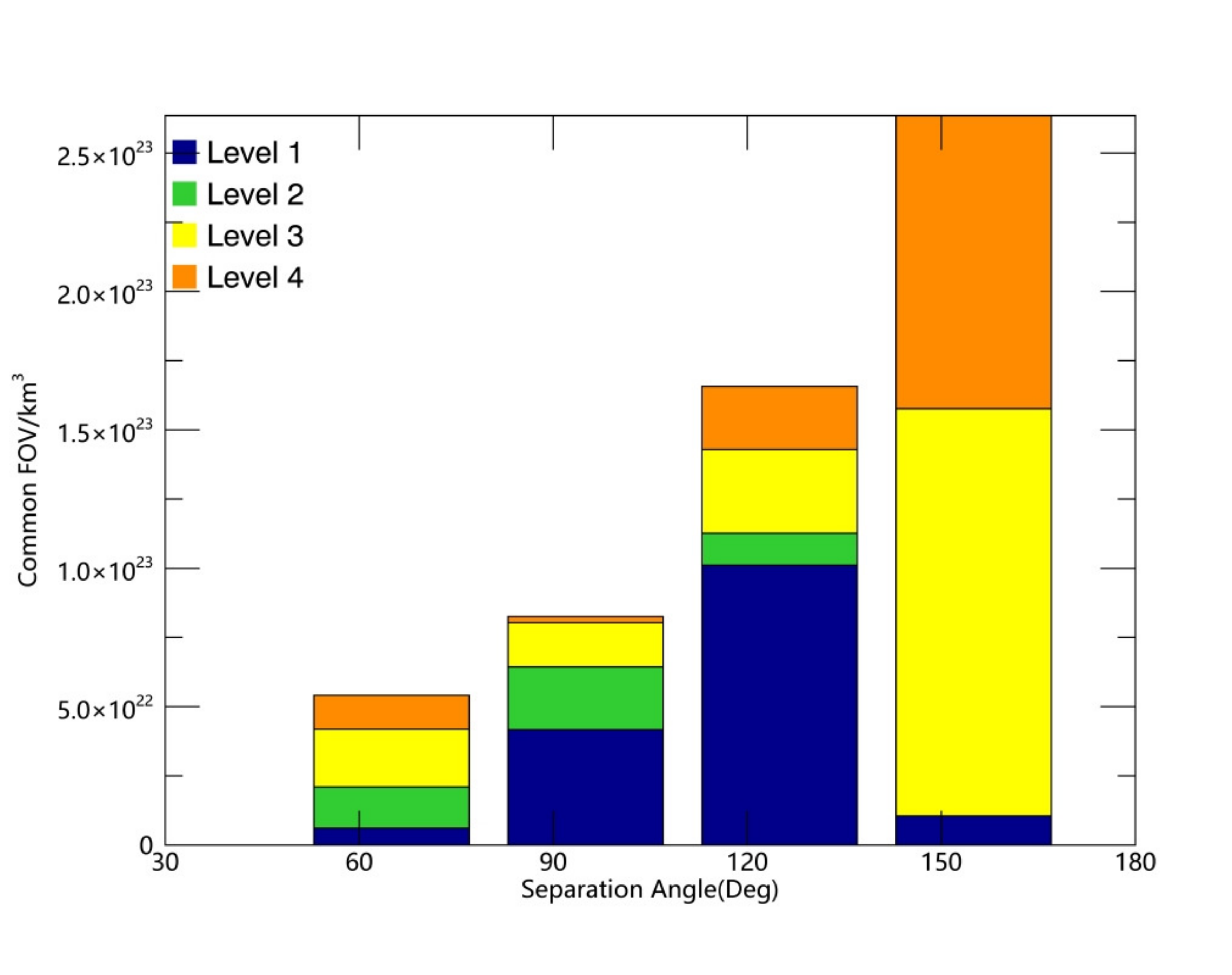}
		}
		\caption{Panel (a): the percentages of the performance levels of the reconstruction for four separation angles. Panel (b): The distribution of the estimated volumes occupied by the four levels.}
		\label{fig:barplot}
	\end{figure*}

	
	With the spatial distribution of reconstructed blobs at four separation angles, we can calculate the percentage of each performance level for the reconstruction and estimate the volume occupied by each level (see Figure \ref{fig:barplot}). In the cases of $60^{\circ}$ and $150^{\circ}$, the area suitable for the reconstruction is small, due to the small \textcolor[rgb]{0,0,0}{volume of the }common FOV and the low proportion, respectively. In the case of $90^{\circ}$, the percentage of Level 1 and Level 2 is the largest, reaching to 77.9\%. This means most region in the case of $90^{\circ}$ is suitable for the reconstruction of structures with a medium or high density, but the total volume of these two performance levels is clearly smaller than that of the $120^{\circ}$ case. Particularly, when the angle of the two spacecraft is $120^{\circ}$, the percentage and the estimated volume of Level 1 are the largest among four cases. Therefore, it could be concluded that the case of $120^{\circ}$ separation angle, out of the four investigated angles, is the best choice for our dual-viewpoint reconstruction. 
	
	\section{Discussion and summary}

	To explore the potential capacity of multi-viewpoint observations from the Solar Ring mission, we apply the CORAR method on reconstructing the blob-like structures throughout the common FOV of dual spacecrafts with different separation angles. The reconstruction results recommend the operation scheme with a dual-viewpoint angle of $120^\circ$ for the instrument arrangement of the Solar Ring mission. With this separation angle, there is the largest region for transients to be well reconstructed. The separation angle of $90^{\circ}$ is next to the case of $120^\circ$ in performance, and is more suitable for the reconstruction of median- or high-density solar structures than low-density transients.

	\begin{figure*}[htb]
		\centering
		\includegraphics[height=0.5\columnwidth,width=0.88\columnwidth]{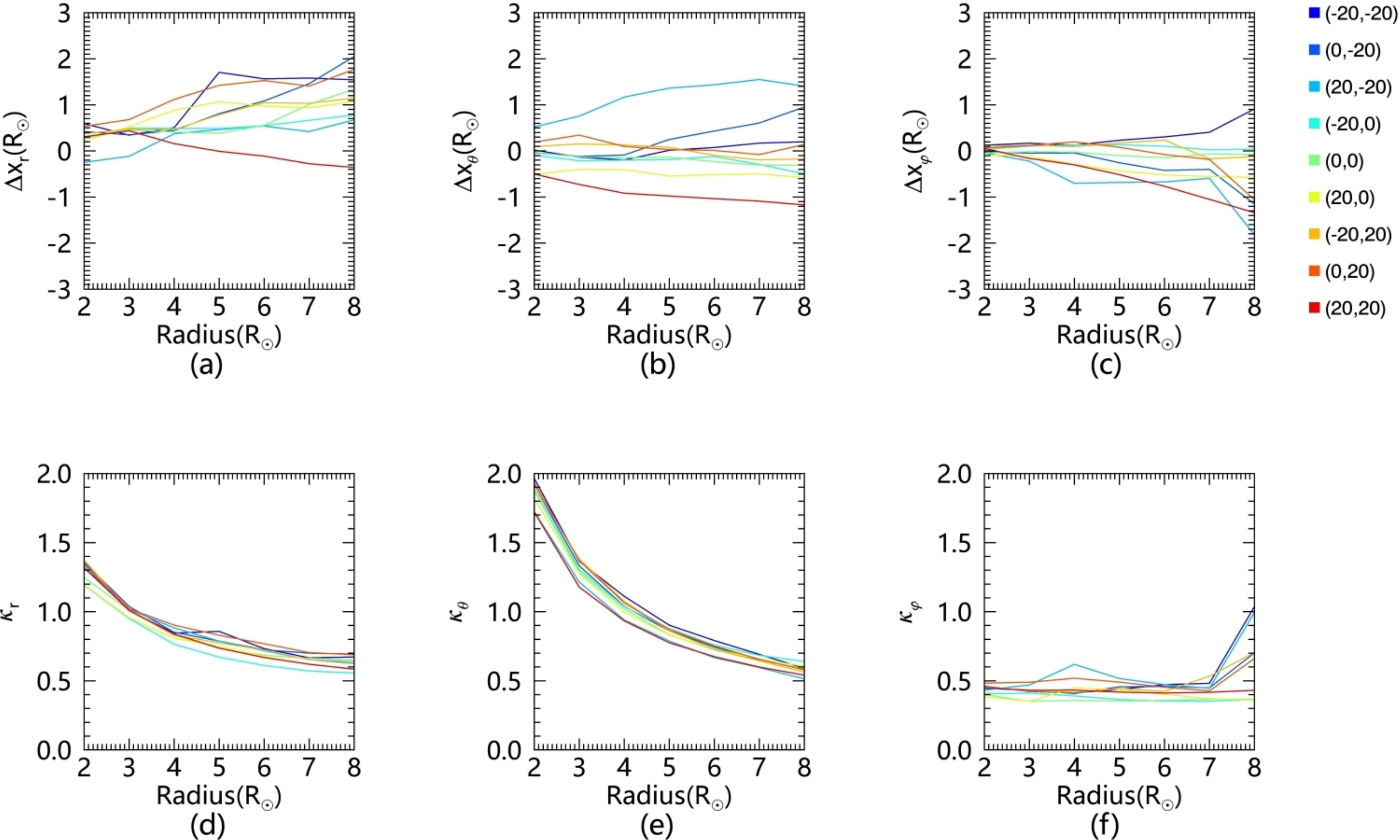}
		\caption{\textcolor[rgb]{0,0,0}{the profiles of criteria parameters PD(panel a-c) and ER(panel d-f) in three dimensions as the function of blob radius $R_0$ when the separation angle is $120^\circ$. The blob HEE coordinates(longitude,latitude) are labelled with different colors, and the distance of blobs from the Sun is 60$R_\odot$. The blob central density is 3000$cm^{-3}$.}}
		\label{fig:size}
	\end{figure*}
	
	Note that our analysis is based on the synthetic images with simple blob-like transients, \textcolor[rgb]{0,0,0}{giving ideal estimates of the reconstruction performance. In reality, the blob geometry is more complex. For the spherical blob model, the reconstruction may vary with different scale size and velocity. The initial size of blobs reported by \citet{N.R.Sheeley1997} at 3-4$R_\odot$ from the Sun is about 1$R_\odot$(radial)$\times$0.1$R_\odot$(transverse), while \citet{Sanchez-Diaz2017} found a typical blob size of 12$R_\odot \times$5$R_\odot$ at the postion of 30$R_\odot$. Considering the expansion of blobs as propagating from the Sun, the radius of blobs may vary in $10^0$-$10^1R_\odot$ in the heliosphere. Figure \ref{fig:size} show the profiles of criteria parameters(Eq.\ref{eq:1}, Eq.\ref{eq:2}) as the function of blob radius $R_0$ with different blob positions at 60$R_\odot$. The blob density is 3000$cm^{-3}$. In three dimensions, PD satisfies the threshold condition, with the absolute value increasing with blob radius. PD in distance is mostly larger than 0 and has an increasing tendency, possibly because the rear-edge patterns are fainter for larger blobs with the same central density. Meanwhile, it is obvious that PD in latitude tends to be positive for southern blobs and negative for northern blobs, and PD in longitude tends to be positive for eastern blobs and negative for western blobs. It means that reconstructed blobs with larger radius far from the Sun-Earth connecting line tend to shift toward larger distance and the central-line area. This error should be seriously considered when reconstructing large-scale transients like CMEs in future. From panel (d)-(f), ER in distance and latitude is apparently decreasing with radius, while in longitude it is mainly between 0.35-0.5 except for blobs away from the Sun-Earth line with extremely large radius. This range of ER is mostly smaller than half of the threshold, reflecting that this method needs improvement for reconstructing blobs completely in longitude. The decrease in distance and latitude is possibly due to the sampling scale for cc calculation: when the scale length of blobs is smaller than that of the sampling area, the spatial size of the reconstructed structure will be close to the latter in the dimensions of sampling. Therefore, for blobs with smaller radius, reducing the sampling size or increasing the grid density should be taken into consideration; otherwise, the reconstruction performance will get worse.} 

	\begin{figure*}[htb]
		\centering
		\includegraphics[height=0.5\columnwidth,width=0.88\columnwidth]{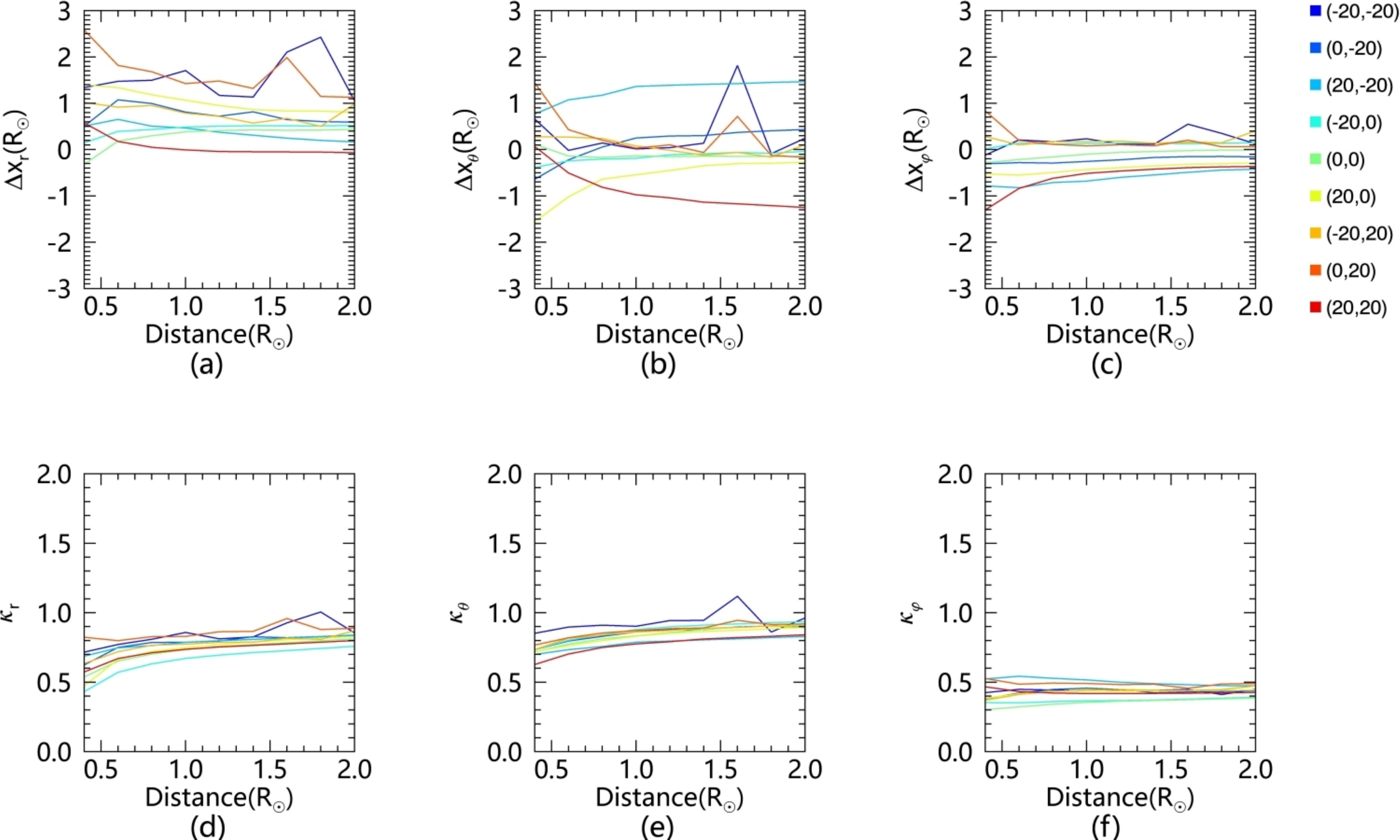}
		\caption{\textcolor[rgb]{0,0,0}{the profiles of criteria parameters PD(panel a-c) and ER(panel d-f) in three dimensions as the function of moving distance of blobs when the separation angle is $120^\circ$. The blob HEE coordinates(longitude,latitude) are labelled with different colors, and the distance of blobs from the Sun is 60$R_\odot$. The blob central density is 3000$cm^{-3}$.}}
		\label{fig:diff}
	\end{figure*}

	\textcolor[rgb]{0,0,0}{The velocity of blobs may vary from 150km/s to less than 500km/s\citep{N.R.Sheeley1997,Sheeley2009,Sheeley2010,Lopez-Portela2018} in the range of 5-50$R_\odot$ from the Sun, which represents 0.5-1.7$R_\odot$ movement in distance within 40 minutes, the time interval between two HI-1 images. To study the influence of various blob speeds on the reconstruction, the profiles of criteria parameters(Eq.\ref{eq:1}, Eq.\ref{eq:2}) as the function of moving distance, which is the radial distance of the blob movement between two observations, are plotted. In Figure \ref{fig:diff}, both PD and ER satisfies the threshold for the selected blob cases. Note that there is deviation in PD profiles between low- and high-velocity limits, possibly due to the running-difference process. With low velocity, the pattern of blobs on images may overlap and disappear in the running-difference counterparts, resulting in the incomplete reconstruction of blobs. This is also the reason for the decrease in the ER profiles as the moving distance decreases in distance and latitude. In longitude, ER for most blob cases are around 0.5, which is similar to the ER profiles in longitude in Figure \ref{fig:size}(f). In summary, the overall reconstruction performance may not change greatly with various blob velocity, while low propagation speed leading to incomplete running-difference patterns can influence the completeness of reconstruction.} 

	In reality, small transients like blobs with a density of up to $10^5cm^{-3}$ are rare in the interplantery space. To capture the patterns of low-density structures, the separation angle may need to be larger to get a vaster common space near the Thomson surfaces. Moreover, for more complicated transients like CMEs, the separation angle larger than $120^{\circ}$ may be more suitable, because two-dimensional patterns based on Thomson scattering integral along LOS from two viewpoints can be completely distinct, except the opening angle is close to $0^\circ$ or $180^\circ$. Thus, the separation angle of $150^\circ$ might be also useful for the reconstruction of large-scale solar wind transients. This will be studied further in the future work.

\begin{figure*}[htb]
		\centering
		\subfigure[]{
			\includegraphics[height=0.39\columnwidth,width=0.44\columnwidth]{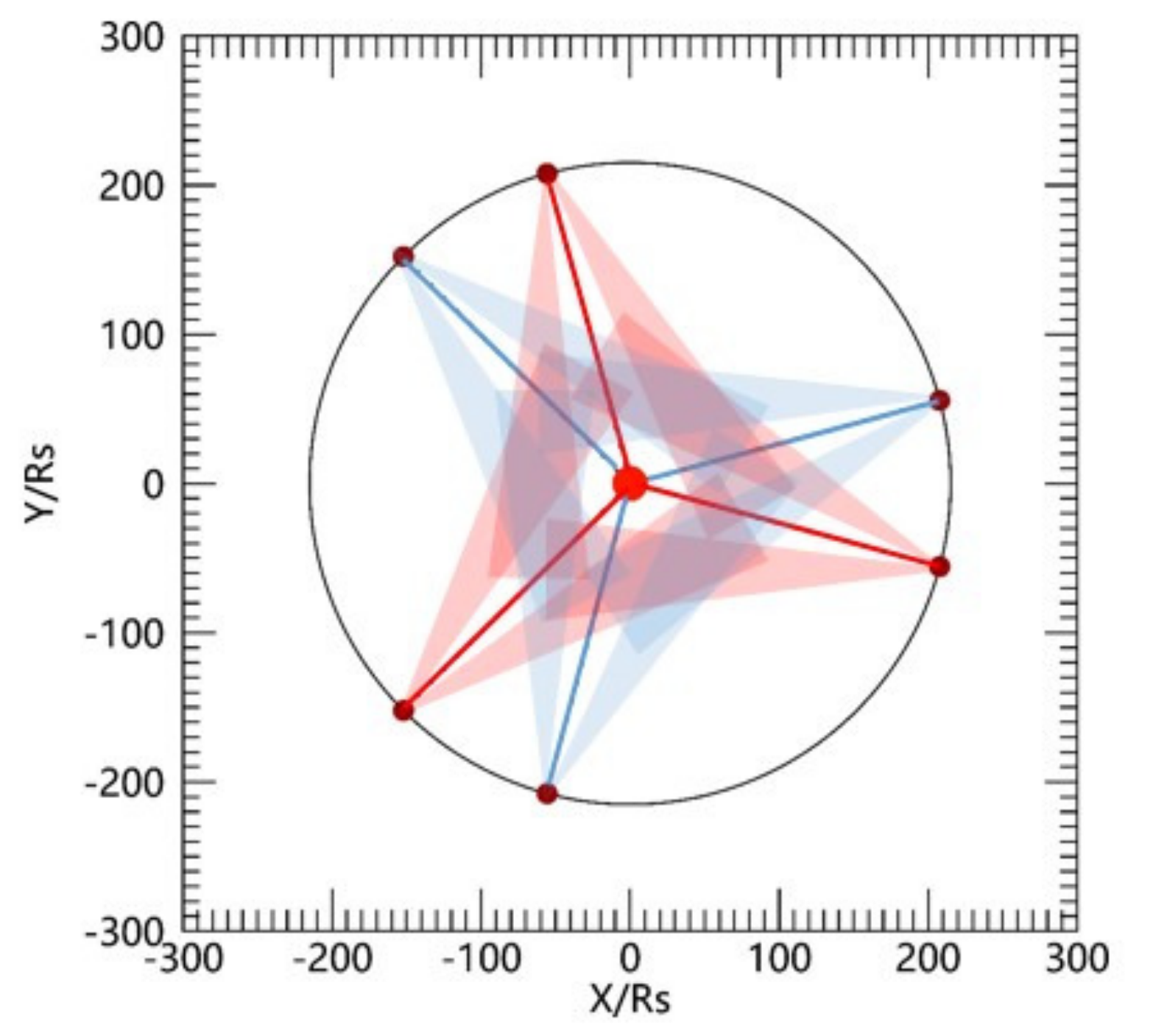}
		}
		\subfigure[]{
			\includegraphics[height=0.39\columnwidth,width=0.44\columnwidth]{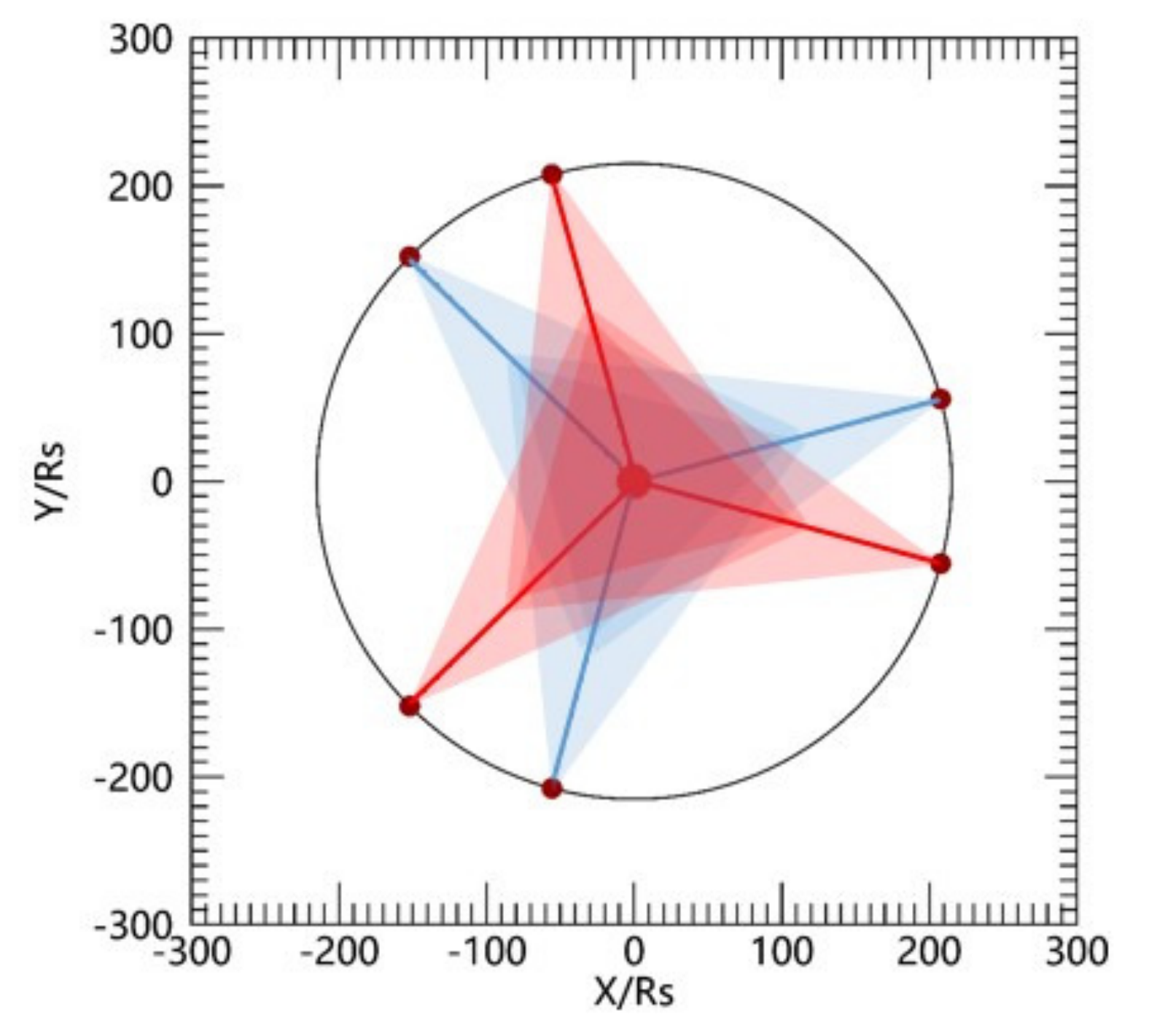}
		}
		\caption{Panel (a): The spacecraft scheme of heliospheric imagers for the Solar Ring mission. Panel (b): The scheme of wide-angle coronagraphs. The cameras in the same color are on board the spacecraft with an separation angle of $120^\circ$.} 
		\label{fig:6}
	\end{figure*}

	Based on the above analysis, we propose the possible arrangements of white-light imagers on board the six spacecraft in the Solar Ring mission\textcolor[rgb]{0,0,0}{(see Figure \ref{fig:2-1})}, as shown in Figure \ref{fig:6}. In the scheme, two heliospheric imagers are on board each spacecraft to observe solar wind simultaneously from two perspectives, or one extra-wide angle coronagraph watching corona and heliosphere on the both sides of the Sun. Such arrangements make possible all $90^{\circ}$, $120^{\circ}$ and $150^{\circ}$ dual-viewpoint observations.
	


	\section*{Acknowledgements}
	The SECCHI data presented in the paper were produced by a consortium of NRL (USA), RAL (UK), LMSAL (USA), GSFC (USA), MPS (Germany), CSL (Belgium), IOTA (France), IAS (France) and obtained from STEREO Science Center (https://stereo-ssc.nascom.nasa.gov/data/ins\_data/). We acknowledge the use of them. This work was supported by the Strategic Priority Program of CAS (Grant Nos.XDA15017300 and XDB41000000), the National Natural Science Foundation of China (Grant Nos.41842037, 41774178, 41761134088 and 41750110481) and the fundamental research funds for the central universities (WK2080000077). 
	
	\vspace{1\baselineskip}
	\noindent \textbf{Conflict of interest} \quad The authors declare that they have no conflict of interest.

	
	
	
	
	\bibliographystyle{model2-names}
	\bibliography{article}
	
	
	
	
	
	
	
\end{document}